\documentclass[%
 aip,
 cha,
 amsmath,amssymb,
preprint,%
]{revtex4-1}
\usepackage{amsmath}
\usepackage{mathrsfs}

\usepackage{graphicx}
\usepackage{dcolumn}
\usepackage{bm}

\newtheorem{prop}{Proposition}

\allowdisplaybreaks

\begin{document}

\preprint{AIP/123-QED}

\title[]{Supersymmetric Reciprocal Transformation and Its Applications}

\author{Q.~P.~Liu}
 \affiliation{Department of Mathematics, China University of Mining and Technology,\\
Beijing 100083, P. R. China}

\author{Ziemowit Popowicz}%
\affiliation{Institute of Theoretical Physics, University of Wroc{\l}aw, pl.
M. Borna 9,50-205, Wroc{\l}aw, Poland}%

\author{Kai Tian}
\affiliation{LSEC, ICMSEC, Academy of Mathematics and Systems Science,\\
Chinese Academy of Sciences, Beijing 100190, P. R. China}%

\date{\today}

\begin{abstract}
The supersymmetric analog of the reciprocal transformation is introduced. This is
used to establish a transformation between one of the supersymmetric Harry Dym equations
and the supersymmetric modified Korteweg-de Vries equation. The
reciprocal transformation,  as a B\"{a}cklund-type transformation between these two
equations, is adopted to construct a recursion operator of the supersymmetric Harry Dym equation.
By proper factorization of the recursion operator, a bi-Hamiltonian structure is found for
the supersymmetric Harry Dym equation. Furthermore, a supersymmetric Kawamoto equation is
proposed and is associated to the supersymmetric Sawada-Kotera equation.  The recursion operator and odd bi-Hamiltonian
structure of the supersymmetric Kawamoto equation are also constructed.
\end{abstract}


\maketitle


\section{Introduction}
The Miura transformation linking the Korteweg-de Vries (KdV) and the
modified KdV (MKdV) equations played a key role in the development
of the Inverse Scattering Method as a technique to solve integrable
nonlinear partial differential equations. On the other side, the gauge
type of transformations among the associated linear problems may be
used to generate transformations between the corresponding nonlinear problems.
Typically the gauge  transformations of the scattering
operator may either change the explicit form of the linear problem -
corresponding to different  nonlinear equations - or they will keep
the linear problem invariant. The first case encodes the Miura
transformation while the latter case represents auto-B\"{a}cklund
transformation.  Moreover the reciprocal transformation, in
conjunction with the gauge transformation also plays a key role in
links between different scattering problems. It was demonstrated in
Ref.\;\onlinecite{rogers} that AKNS and WKI scattering schemes are linked,
when the transformations of the independent variables are taken
into account. Furthermore in the context of soliton theory, the Harry Dym
(HD) hierarchy \cite{calogero,ibragimov}, known to be
invariant under a reciprocal transformation,  is also connected with
the KdV hierarchy. The Camassa-Holm equation is connected via the
reciprocal link with the first negative flow of the KdV hierarchy \cite{fuch},
while the  Sawada-Kotera (SK) and Kaup-Kupershmidt (KK) equations are linked
by reciprocal transformation with the Kawamoto equation \cite{kawa}.

The main purpose of the present paper is to describe how we should modify the scenario
of the reciprocal transformation to  the supersymmetric equations. Our aim is to establish a proper
 transformation between the supersymmetric HD and
MKdV equations. Apart from its own interest, we are mainly concerned with revealing the integrability properties,
such as recursion operators and bi-Hamiltonian structures, for the supersymmetric HD equation. The fifth order equations also will be
studied.

Our approach is  motivated by recent interest to  the supersymmetric nonlinear
partial differential equations. These equations have long history and appeared
almost in parallel to the usage of the supersymmetry in the quantum field theory. The first
results, concerned the construction of classical field theories with  fermionic
and bosonic fields depending on time and one space variable, can be found in Refs.\;\onlinecite{chai,sci,oguz,kulish,kup,man}.
In many cases, the addition of fermionic fields does not guarantee that the final
theory becomes supersymmetric invariant. Therefore this method was named as
the fermionic extension in order to distinguish it from the fully supersymmetric
method which was developed later \cite{mat,lab,labe,chaiL,oevel}.

There are many recipes how   classical  models could be embedded  in
fully supersymmetric superspace. The main idea is simple: in order
to get such generalization we should construct the  supermultiplet
which contains  the classical functions. It means that we have to
add to a system of $k$ bosonic equations $kN$ fermions and $k(N-1)$
bosons ($k=1,2,..., N=1,2,...$) in such  a way that they create
superfields. Now working with this supermultiplet we can step by
step  apply  integrable Hamiltonian  methods to our considerations
depending on what we would like to construct. In that way the basic
solitonic equations have been supersymmetrized as for example the
KdV equation, Boussinesq equation, Two-Boson equation, HD equation and recently the SK
equation. All these supersymmetric equations are integrable in the
sense that they possess a Lax representation or a recursion operator or a bi-Hamiltonian
structure and consequently they have infinite number of conserved
densities. Interestingly {\em not all solitonic equations have been
successfully embedded into  the superspace}, as for example we do
not know up to now the supersymmetric version of the Kaup-Kupershmidt (KK) equation.

Due to the large number of the supersymmetric equations it is
reasonable to find the reciprocal link between these  supersymmetric
equations. Our paper concerns with this problem and is arranged as follows.
In the following section we recapitulate  basic notations
and ideas used in the classical reciprocal link between the HD
and MKdV equations. Section \ref{s3}, after
introducing the supersymmetric notation,  deals with the
supersymmetric generalization of the reciprocal  link for  the
supersymmetric HD and MKdV equations. The reciprocal transformation
defined in Section \ref{s3} is applied  to construct the missing recursion operator, and
bi-Hamiltonian structure of the SUSY HD equation in Section \ref{s4}.
Section \ref{s5} presents a new SUSY generalization of Kawamoto equation, its Lax representation and construction
of the supersymmetric reciprocal link to the recently discovered
SUSY SK equation. The recursion operator, and odd bi-Hamiltonian structure
of the SUSY Kawamoto equation is calculated  in Section \ref{s6}.
The last section contains concluding remarks.

\section{Review on the classical case}\label{s2}
The integrability of the Harry Dym equation
\begin{equation}\label{hd}
u_t=-u^3u_{3x}
\end{equation}
follows from its nonstandard Lax representation
\begin{equation}\label{hdlax}
\left[\frac{\partial}{\partial t}+B,\; L\right]=0,
\end{equation}
where
\begin{equation*}
L=u^2\partial_x^2,\quad
B=4(L^{2/3})_{\geq 2}=4u^3\partial_x^3+6u^2u_x\partial_x^2,
\end{equation*}
or from its bi - Hamiltonian  formulation
\begin{equation}
 u_t= u^3 \partial_{x}^3u^3 \frac{\delta}{\delta u} \int {\tt d}x \;u^{-1} = u^2\partial_{x}u^2 \frac{\delta}{\delta u}
\int {\tt d}x\; \frac{u_x^2}{2u}.
\end{equation}
To associate the HD equation \eqref{hd} to the MKdV equation, we recall the Liouville transformation. Namely,
\begin{equation}\label{liou}
\frac{\partial}{\partial x}=s^{-1}\frac{\partial}{\partial y},
\end{equation}
where $s(y,\tau)=u(x,t)$, then the Lax operator $L$ is transformed to
\begin{equation*}
\hat{L}=\partial_y^2+v\partial_y,
\end{equation*}
where $v$ is given by the Cole-Hopf transformation
\begin{equation}\label{cole}
v=-\frac{\partial}{\partial y}\log s.
\end{equation}
It is straightforward to check that the Lax equation
\begin{equation}\label{mkdvlax}
\left[\frac{\partial}{\partial\tau}+\hat{B},\; \hat{L}\right]=0
\end{equation}
implies the MKdV equation
\begin{equation}\label{mkdv}
v_\tau=-v_{3y}+\frac{3}{2}v^2v_y,
\end{equation}
where
\begin{equation*}
\hat{B}=4(\hat{L}^{3/2})_{\geq 1}=4\partial_y^3+6v\partial_y^2+\left(3v_y+\frac{3}{2}v^2\right)\partial_y.
\end{equation*}

Now, the relation between  $\partial_t$ and $\partial_\tau$ is
inferred from formulas \eqref{hdlax} and \eqref{mkdvlax}, which is given by
\begin{eqnarray*}
\frac{\partial}{\partial t}&=&\frac{\partial}{\partial \tau}+\hat{B}\mid_{v=-s_y/s}
-B\mid_{u=s,\;\partial_x=s^{-1}\partial_y}\\
&=&\frac{\partial}{\partial \tau}+\left(s^{-1}s_{2y}-\frac{3}{2}s^{-2}s_y^2\right)\frac{\partial}{\partial y}\\
&=&\frac{\partial}{\partial \tau}+\left(uu_{2x}-\frac{1}{2}u_x^2\right)\frac{\partial}{\partial y}.
\end{eqnarray*}

Thus, we recover the reciprocal link between the HD equation \eqref{hd} and the MKdV equation \eqref{mkdv}, i.e.
\begin{eqnarray*}
\frac{\partial}{\partial x}&=&u^{-1}\frac{\partial}{\partial y},\\
\frac{\partial}{\partial t}&=&\left(uu_{2x}-\frac{1}{2}u_x^2\right)\frac{\partial}{\partial y} +\frac{\partial}{\partial \tau},
\end{eqnarray*}
and the relation between two fields defined by \eqref{cole}. The explicit transformations for the independent variables are given by
\begin{equation*}
y=\int {\tt d}x\;u^{-1} \qquad \tau=t.
\end{equation*}

The evolution of the field $s$ is governed by
\begin{equation}\label{psckdv}
\frac{\partial}{\partial\tau}s=-\frac{\partial}{\partial y}\left(s_{2y}-\frac{3}{2}\frac{s_y^2}{s}\right),
\end{equation}
which, by introducing the potential $s=w_y$, is transformed to the Schwarzian KdV equation
\begin{equation}\label{sckdv}
w_\tau=-\left(w_{3y}-\frac{3}{2}\frac{w_{2y}^2}{w_y}\right)=-w_y\{w,y\},
\end{equation}
where $\{w,y\}$ denotes the Schwarzian derivative of $w$ with respect to $y$.

To obtain the MKdV equation \eqref{mkdv} from the Schwarzian KdV equation \eqref{sckdv}, one only need to take the Cole-Hopf transformation
\begin{equation*}
v=-\frac{\partial}{\partial y}\log w_y.
\end{equation*}

We remark that above construction of the reciprocal transformation heavily relies on the Lax representation.
An alternative is based on the conservation laws. Indeed, the HD equation \eqref{hd} can be reformulated as follows
\begin{equation*}
\frac{\partial}{\partial t}\Big(u^{-1}\Big)=\frac{\partial}{\partial x}\left(uu_{2x}-\frac{1}{2}u_x^2\right).
\end{equation*}
Thus, it is natural to introduce \cite{kawa}
\begin{equation*}
{\tt d}y=u^{-1}{\tt d}x+\left(uu_{2x}-\frac{1}{2}u_x^2\right){\tt d}t,\qquad {\tt d}\tau={\tt d}t
\end{equation*}
which is nothing but the reciprocal transformation discussed. The advantage of this latter approach is that everything follows from the equation {\em only}.

\section{SUSY reciprocal transformation: SUSY Harry Dym case}\label{s3}
In this section, taking one of the SUSY HD equations proposed in Ref.\;\onlinecite{brunelli} as an example,
we exhibit the reciprocal link between the SUSY HD and MKdV equations.

The SUSY HD equation we will consider takes the form \cite{brunelli}
\begin{eqnarray*}
 W_t&=&\frac{1}{16}\Big[8{\cal D}^5(({\cal D}W)^{-1/ 2})-3{\cal D}(W_{xx}W_x({\cal D}W)^{-5/2})\\
&&+\frac{3}{4}({\cal D}W_x)^2W_x({\cal D}W)^{-7/2}-\frac{3}{4}{\cal D}^{-1}\left(({\cal D}W_x)^3({\cal D}W)^{-7/2}\right)\Big].
\end{eqnarray*}
where $W=W(x,\theta,t)$ is a fermionic super field and ${\cal D}=\partial_{\theta}+\theta{\partial_x}$ is the superderivative.

By means of
\[
U=(\mathcal{D}W)^{-1/2},
\]
this equation can be more conveniently written as
\begin{equation}\label{shd}
U_t=\frac{1}{4}U_{3x}U^3-\frac{3}{8}(\mathcal{D}U_{2x})(\mathcal{D}U)U^2,
\end{equation}
which admits the Lax representation
\begin{equation}\label{shklax}
\left[\frac{\partial}{\partial t}-(L_h^{3/2})_{\geq 3},\; L_h\right]=0,
\end{equation}
with the Lax operator $L_h=U\mathcal{D}U\partial_x\mathcal{D}$.

\subsection{The super analogy of Liouville transformation}\label{3-1}

Our aim now is to convert the Lax operator $L_h$ of the SUSY HD equation into that of the SUSY MKdV equation. To this end, we propose the following  super analogy of
Liouville transformation
\begin{equation}\label{sliou}
\mathcal{D}=S^{-1/2}\mathbb{D}~,
\end{equation}
where $S(y,\varrho,\tau)=U(x,\theta,t)$ and $\mathbb{D}$ denotes the transformed superderivative given by
\begin{equation*}
\mathbb{D}=\partial_\varrho+\varrho\partial_y.
\end{equation*}

Through the transformation (\ref{sliou}), we obtain from $L_h$
\begin{equation*}
L_m=\partial_y^2+(\mathbb{D}\Psi)\partial_y+\left(\frac{1}{2}\Psi_y
+\frac{1}{4}\Psi(\mathbb{D}\Psi)\right)\mathbb{D},
\end{equation*}
where $\Psi$ is a fermionic super field related with $S$ by the super Cole-Hopf transformation
\begin{equation}\label{scole}
\Psi=-\mathbb{D}\log S.
\end{equation}

We claim that the operator $L_m$ is the Lax operator of the SUSY MKdV equation. In fact, an easy calculation shows that the Lax equation
\begin{equation}\label{smkdvlax}
\left[\frac{\partial}{\partial\tau}-(L_m^{3/2})_{\geq 1},\; L_m\right]=0
\end{equation}
implies the SUSY MKdV equation
\begin{equation}\label{smkdv}
\Psi_\tau=\frac{1}{4}\Psi_{3y}-\frac{3}{16}(\mathbb{D}\Psi)[\Psi(\mathbb{D}\Psi)]_y.
\end{equation}

The transformation between $\partial_t$ and $\partial_\tau$ is found  by
\begin{eqnarray}\label{stem}
\frac{\partial}{\partial t}&=&\frac{\partial}{\partial\tau}-(L_m^{3/2})_{\geq 1}\mid_{\Psi=-\mathbb{D}\log S}
+(L_h^{3/2})_{\geq 3}\mid_{U=S,\;\mathcal{D}=S^{-1/2}\mathbb{D}}\nonumber\\
&=&\frac{\partial}{\partial\tau}+\frac{1}{8}\left(-2\frac{S_{2y}}{S}+3\frac{S_y^2}{S^2}+\frac{(\mathbb{D}S_y)
(\mathbb{D}S)}{S^2}\right)\frac{\partial}{\partial y}\nonumber\\
&&+\frac{1}{16}\left(-2\frac{(\mathbb{D}S_{2y})}{S} +5\frac{(\mathbb{D}S_y)S_y}{S^2}+3\frac{(\mathbb{D}S)S_{2y}}{S^2} -6\frac{(\mathbb{D}S)S_y^2}{S^3}\right)\mathbb{D}\nonumber\\
&=&\frac{\partial}{\partial\tau}+\left(-\frac{1}{4}U_{2x}U+\frac{1}{8}U_x^2
+\frac{1}{4}(\mathcal{D}U_x)(\mathcal{D}U)\right)\frac{\partial}{\partial y}\nonumber\\
&&+\left(-\frac{1}{8}(\mathcal{D}U_{2x})U^{3/2}\right)\mathbb{D}.
\end{eqnarray}

Therefore, we succeeded in constructing between the SUSY HD equation (\ref{shd}) and the SUSY MKdV equation (\ref{smkdv}) a reciprocal transformation
\begin{equation*}
(x,\theta,t,U)\rightarrow(y,\varrho,\tau,\Psi)~,
\end{equation*}
which is given by formulas (\ref{sliou}), (\ref{scole}) and (\ref{stem}).

In terms of the  new super space-time $(y,\varrho,\tau)$, the bosonic field $S$ satisfies
\begin{equation}\label{spsckdv}
S_\tau=\frac{1}{16}\left(4S_{3y}-12\frac{S_{2y}S_y}{S}+6\frac{S_y^3}{S^2}-3\frac{(\mathbb{D}S_y)(\mathbb{D}S)S_y}{S^2}\right),
\end{equation}
from which we obtain, through the transformation $S=(\mathbb{D}\Lambda)^{-2}$, the SUSY Schwarzian KdV equation \cite{mathieu}
\begin{equation}\label{ssckdv}
\Lambda_\tau=\frac{1}{4}\left(\Lambda_{3y}-3\frac{\Lambda_{2y}(\mathbb{D}\Lambda_y)}{(\mathbb{D}\Lambda)}\right).
\end{equation}
The link between the SUSY Schwarzian KdV equation (\ref{ssckdv}) and the SUSY MKdV equation (\ref{smkdv}) is supplied by the super Cole-Hopf transformation $\Psi=2\mathbb{D}\log (\mathbb{D}\Lambda)$.

\subsection{The superconformal transformation}
Relying on the linear spectral problem, we constructed the reciprocal link between the SUSY HD equation and the SUSY MKdV equation  in last subsection \ref{3-1}. It would be nice if the reciprocal transformation can be established without any knowledge of linear problems. Next we will show that
 this is indeed the case: the reciprocal transformation is the result of the equation itself.  To this end, we  need the superconformal transformation, which is the super diffeomorphism such that the superderivative transforms covariantly \cite{mathieu, friedan, arvis}. Namely, $\mathcal{D}=G\mathbb{D}$, where $G$ is a bosonic super field.

\begin{prop}\label{prop1}
Let
\begin{equation}\label{conserv}
\frac{\partial G}{\partial t}={\cal D}\Xi
\end{equation}
be a conservation law, and suppose that a potential $W$ can be introduced by
\begin{equation}\label{pont}
{\cal D}W=2\Xi G,
\end{equation}
then a superconformal transformation
may be defined consistently.
\end{prop}


\noindent{\bf Proof: }
First, we consider the following change of variables
\begin{equation}\label{susyconformal}
(x,\theta,t)\rightarrow(y,\varrho,\tau)=(y(x,\theta,t),\varrho(x,\theta,t),t)
\end{equation}
where $(y, \varrho)$ is the new super spatial variable and $\tau$ is the new temporal variable. Next we use notation
$\mathbb{D}=\frac{\partial}{\partial \varrho}+\varrho\frac{\partial}{\partial y}$ as our new superderivative.

To ensure that (\ref{susyconformal}) is a superconformal transformation, we impose
\begin{equation}\label{susyconformal-1}
\mathcal{D}=G\mathbb{D},
\end{equation}
and by  a direct calculation we have
\begin{eqnarray*}
&&\frac{\partial}{\partial x}=\left[\frac{\partial y}{\partial x} -\left(\frac{\partial\varrho}{\partial x}\right)\varrho\right]\frac{\partial}{\partial y} +\frac{\partial\varrho}{\partial x}\mathbb{D}
\equiv S\frac{\partial}{\partial y}+\Gamma\mathbb{D},\\
&&\frac{\partial}{\partial \theta}=\left[\frac{\partial y}{\partial\theta} -\left(\frac{\partial\varrho}{\partial \theta}\right)\varrho\right]\frac{\partial}{\partial y} +\frac{\partial\varrho}{\partial \theta}\mathbb{D}
\equiv \Lambda\frac{\partial}{\partial y}+T\mathbb{D},\\
&&\frac{\partial}{\partial t}=\left[\frac{\partial y}{\partial t}-\left(\frac{\partial\varrho}{\partial t}\right)\varrho\right]\frac{\partial}{\partial y} +\frac{\partial\varrho}{\partial t}\mathbb{D}+\frac{\partial}{\partial\tau}
\equiv \hat{W}\frac{\partial}{\partial y}+\hat{\Xi}\mathbb{D}+\frac{\partial}{\partial\tau}.
\end{eqnarray*}
Naturally, all the coefficients $S$, $T$, $\hat{W}$, $\Gamma$, $\Lambda$ and $\hat{\Xi}$ have to be selected consistently such that the
compatibility conditions
\begin{align*}
&\frac{\partial}{\partial\theta}\Gamma=\frac{\partial}{\partial x}T,\quad
&&\frac{\partial}{\partial\theta}\hat{\Xi}=\frac{\partial}{\partial t}T,\quad
&&\frac{\partial}{\partial x}\hat{\Xi}=\frac{\partial}{\partial t}\Gamma,\\
&\frac{\partial}{\partial\theta}S=\frac{\partial}{\partial x}\Lambda+2\Gamma T,\quad
&&\frac{\partial}{\partial t}\Lambda=\frac{\partial}{\partial\theta}\hat{W}-2\hat{\Xi} T,\quad
&&\frac{\partial}{\partial t}S=\frac{\partial}{\partial x}\hat{W}+2\hat{\Xi}\Gamma,
\end{align*}
hold.

We notice that (\ref{susyconformal-1}) implies
\[
\frac{\partial}{\partial x}=\mathcal{D}^2=G\mathbb{D}G\mathbb{D}=G^2\frac{\partial}{\partial y}+(\mathcal{D}G)\mathbb{D},
\]
and also we have
\begin{eqnarray*}
\frac{\partial}{\partial\theta}&=&\mathcal{D}-\theta\frac{\partial}{\partial x}=G\mathbb{D}-\theta\left(G^2\frac{\partial}{\partial y}+(\mathcal{D}G)\mathbb{D}\right)\\
&=&-\theta G^2\frac{\partial}{\partial y}+\Big(G-\theta(\mathcal{D}G)\Big)\mathbb{D}.
\end{eqnarray*}
These last two equations suggest the following identifications
\[
S=G^2,\qquad  \Gamma=(\mathcal{D}G),\qquad \Lambda=-\theta G^2,\qquad T=G-\theta(\mathcal{D}G),
\]
furthermore, we ask for
\[
\hat{W}=W,\qquad \hat{\Xi}=\Xi.
\]
Then, a straightforward calculation shows that all the compatibility conditions are satisfied and thus, the proposition  is proved.
\hfill$\square$

Let us now consider the SUSY HD equation (\ref{shd}). In order to
take advantage of  the Proposition \ref{prop1},  we notice that the SUSY HD
equation (\ref{shd}) has the following conservation law
\begin{equation}\label{conser}
\frac{\partial}{\partial t}\Big(U^{-1/2}\Big)=\mathcal{D}\left(-\frac{1}{8}(\mathcal{D}U_{2x})U^{3/2}\right),
\end{equation}
which leads to
\[
G=U^{-1/2},\qquad \Xi=-\frac{1}{8}(\mathcal{D}U_{2x})U^{3/2},
\]
also,
\[
2\Xi G=-\frac{1}{4}(\mathcal{D}U_{2x})U=\mathcal{D}\left(-\frac{1}{4}U_{2x}U+\frac{1}{8}U_x^2+\frac{1}{4}(\mathcal{D}U_x)(\mathcal{D}U)\right),
\]
hence, we choose
\begin{equation*}
\mathcal{D}=U^{-1/2}\mathbb{D},
\end{equation*}
which coincides with the super analogy of Liouville transformation (\ref{sliou}) and
\[
W=-\frac{1}{4}U_{2x}U+\frac{1}{8}U_x^2+\frac{1}{4}(\mathcal{D}U_x)(\mathcal{D}U).
\]
 According to our proposition, the relation between
$\partial_t$ and $\partial_\tau$ is given by
\[
\frac{\partial}{\partial
t}=\left(-\frac{1}{4}U_{2x}U+\frac{1}{8}U_x^2+\frac{1}{4}(\mathcal{D}U_x)(\mathcal{D}U)\right)\frac{\partial}{\partial
y}+\left(-\frac{1}{8}(\mathcal{D}U_{2x})U^{3/2}\right)\mathbb{D}+\frac{\partial}{\partial\tau},
\]
which of course recovers the relation (\ref{stem}), as is expected.

\section{Bi-Hamiltonian structure of the SUSY HD equation}\label{s4}
In classical theory, various interesting properties of the HD equation could be constructed with the help of the reciprocal transformation. Indeed, based on the
general results (Theorem 1 and Theorem 2 in Ref.\;\onlinecite{fokfu}),  one could deduce its hereditary recursion operator from that of the equation \eqref{psckdv}. Let's recall this briefly.\cite{fuch_car}

The recursion operator of the equation \eqref{psckdv} reads
\begin{equation*}
R(s)=\partial_ys\partial_y^{-1}s\partial_ys^{-1}\partial_ys^{-1},
\end{equation*}
which is easily obtained from that of the MKdV equation \eqref{mkdv} through the Cole-Hopf transformation \eqref{cole}.
The reciprocal transformation between equations \eqref{hd} and \eqref{psckdv} may be interpreted as a B\"{a}cklund transformation,
\begin{equation*}
B(u,s)=u(x,t)-s(\partial_x^{-1}u^{-1},t)=0,
\end{equation*}
which can be used to construct the recursion operator for the HD equation from that of equation \eqref{psckdv}. To this end, one has to calculate
the Fr\'{e}chet derivatives of $B(u,s)$ in directions $u$ and $s$ respectively, which are given by
\begin{equation*}
B_s=-1\quad\mbox{and}\quad B_u=u\partial_xu\partial_x^{-1}u^{-2}.
\end{equation*}

The recursion operator of the HD equation \eqref{hd} is given by
\begin{eqnarray*}
R(u)&=&B_u^{-1}B_sR(s)\mid_{s=u,\;\partial_y=u\partial_x}B_s^{-1}B_u\\
&=&u^2\partial_xu^{-1}\partial_x^{-1}u^{-1}\cdot u\partial_xu\partial_x^{-1}u\partial_x^2u^{-1}\cdot u\partial_xu\partial_x^{-1}u^{-2}\\
&=&u^3\partial_x^3u\partial_x^{-1}u^{-2}.
\end{eqnarray*}
A simple decomposition
\begin{equation*}
R(u)=u^3\partial_x^3u^3\cdot \Big(u^{2}\partial_xu^{-2}\Big)^{-1},
\end{equation*}
yields two well-known Hamiltonian operators of the HD equation \eqref{hd} (see Ref.\;\onlinecite{kono} for example).

Not much is known for the SUSY HD equation \eqref{shd}. In contrast, the SUSY KdV (or MKdV) equation has been
investigated extensively and thoroughly, and many properties have been established, including those common properties shared by both classical and supersymmetric systems, and some novel properties owned by supersymmetric systems. Now a bridge between the SUSY HD and MKdV equations has
been builded, thus it is natural to expect that this result will aid us to reveal more properties for the SUSY HD equation and understand it better.

We will show that, as  in classical case, the recursion operator of the SUSY HD equation \eqref{shd} can be constructed through the reciprocal transformation. Furthermore, combining our recursion operator with  a simple Hamiltonian operator of the SUSY HD equation obtained through the $r$-matrix approach, we find its bi-Hamiltonian structure. To reach these results,
the bi-Hamiltonian structures of the SUSY KdV, MKdV and Schwarzian KdV equations have to be recalled and evaluated, which will be done in the following subsection.

\subsection{Bi-Hamiltonian structures of the SUSY KdV, MKdV and Schwarzian KdV equations}
The SUSY KdV equation reads
\begin{equation}\label{skdv}
\Phi_\tau=\Phi_{3y}+3[\Phi(\mathbb{D}\Phi)]_y,
\end{equation}
which is a bi-Hamiltonian system \cite{oevel}, i.e.
\begin{eqnarray}
J(\Phi)\Phi_\tau&=&\frac{\delta }{\delta\Phi}\int\verb"d"y\verb"d"\varrho\;\Big(\Phi(\mathbb{D}\Phi)^2-\frac{1}{2}\Phi_y(\mathbb{D}\Phi_y)\Big),\\
\Phi_\tau&=&P(\Phi)\frac{\delta}{\delta\Phi} \int\verb"d"y\verb"d"\varrho\;\frac{1}{2}\Phi(\mathbb{D}\Phi),\
\end{eqnarray}
where the symplectic operator $J(\Phi)$ and the Hamiltonian operator $P(\Phi)$ are given by
\[
J(\Phi)=\partial_y^{-1}(\mathbb{D}^3+\Phi)\partial_y^{-1},\qquad
P(\Phi)=\partial_y^2\mathbb{D}+2\partial_y\Phi+2\Phi\partial_y+\mathbb{D}\Phi\mathbb{D}.
\]

Under the super Miura transformation
\[
\Phi=\Psi_y-\Psi(\mathbb{D}\Psi),
\]
the SUSY KdV equation is transformed to the SUSY MKdV equation
\begin{equation}\label{smkdv2}
\Psi_\tau=\Psi_{3y}-3(\mathbb{D}\Psi)[\Psi(\mathbb{D}\Psi)]_y,
\end{equation}
which is equivalent to the equation \eqref{smkdv} up to the scaling $\tau\rightarrow\tau/4$, $\Psi\rightarrow\Psi/2$.
The bi-Hamiltonian structure of this equation is given by
\begin{eqnarray*}
J(\Psi)\Psi_\tau&=&\frac{\delta}{\delta\Psi}\int\verb"d"y\verb"d"\varrho\;\Big(-{5\over6}\Psi_{2y}\Psi_y\Psi(\mathbb{D}\Psi)
-{1\over2}\Psi(\mathbb{D}\Psi_{4y})+{5\over3}\Psi(\mathbb{D}\Psi_{2y})(\mathbb{D}\Psi)^2
-\Psi(\mathbb{D}\Psi)^5\Big),\\
\Psi_\tau&=&-\mathbb{D}\frac{\delta}{\delta\Psi}\int\verb"d"y\verb"d"\varrho\;\frac{1}{2}\Big(\Psi_y(\mathbb{D}\Psi_y) -\Psi(\mathbb{D}\Psi_y)(\mathbb{D}\Psi)+\Psi(\mathbb{D}\Psi)^3\Big),
\end{eqnarray*}
where $J(\Psi)$ and $-\mathbb{D}$ are canonically related to $J(\Phi)$  and $P(\Phi)$, respectively. Explicitly
\begin{equation*}
J(\Psi)=4\mathbb{D}\Psi\mathbb{D}^{-1}\Psi\mathbb{D}
-\Big(\mathbb{D}-\Psi\mathbb{D}^{-1}\Psi-2\mathbb{D}\Psi\partial_y^{-1}\Psi\Big)
\mathbb{D}\Big(\mathbb{D}-\Psi\mathbb{D}^{-1}\Psi-2\Psi\partial_y^{-1}\Psi\mathbb{D}\Big).
\end{equation*}

By means of the super Cole-Hopf transformation $\Psi=\mathbb{D}\log V$, one obtains the following equation
\begin{equation}\label{schw}
V_\tau=\mathbb{D}\left((\mathbb{D}V_{2y})-3\frac{V_y(\mathbb{D}V_y)}{V}\right)
\end{equation}
from which we arrive at the SUSY Schwarzian KdV equation \eqref{ssckdv} by taking the potential $V=(\mathbb{D}\Lambda)$
and a time scaling. The equation \eqref{schw} is formulated as a Hamiltonian system by either $J(V)$ or $P(V)$, namely
\begin{eqnarray*}
J(V)V_\tau=\frac{\delta}{\delta V}H_1(V),\quad V_\tau=P(V)\frac{\delta}{\delta V}H_0(V),
\end{eqnarray*}
where
\begin{eqnarray*}
J(V)&=&V^{-1}\mathbb{D}V^{-1}\mathbb{D}V^{-1}\mathbb{D}V^2\partial_y^{-1}V\mathbb{D}V^{-1}\mathbb{D}V^{-1}\mathbb{D}V
\partial_y^{-1}V^2\mathbb{D}V^{-1}\mathbb{D}V^{-1}\mathbb{D}V^{-1},\\
P(V)&=&V\mathbb{D}^{-1}V
\end{eqnarray*}
are obtained from the relations
\begin{eqnarray*}
J(V)=\Psi_V^{\dag}J(\Psi)\mid_{\Psi=\mathbb{D}\log V}\Psi_V\qquad\mbox{and}\qquad
-\mathbb{D}=(\Psi_V)^{-1}P(V)(\Psi_V^{\dag})^{-1}.
\end{eqnarray*}
($\Psi_V$ denotes the Fr\'{e}chet derivative of $\Psi=\mathbb{D}\log V$ in the direction $V$). The Hamiltonians, which can be easily deduced from
those of the SUSY MKdV equation, are
\begin{eqnarray*}
H_0(V)=\int\verb"d"y\verb"d"\varrho&&\; \frac{1}{2}\Big\{(\mathbb{D}V_y)V_{2y}V^{-2}-(\mathbb{D}V_y)V_y^2V^{-3}-2(\mathbb{D}V)V_{2y}V_yV^{-3}
+3(\mathbb{D}V)V_y^3V^{-4}\Big\} \\
H_1(V)=\int\verb"d"y\verb"d"\varrho&&\;\Big\{-{5\over6}(\mathbb{D}V_{2y})(\mathbb{D}V_{y})(\mathbb{D}V)V_yV^{-4}-{1\over2}(\mathbb{D}V)V_{5y}V^{-2}
+{5\over2}(\mathbb{D}V)V_{4y}V_yV^{-3}\\
&&+5(\mathbb{D}V)V_{3y}V_{2y}V^{-3}-{25\over3}(\mathbb{D}V)V_{3y}V_y^2V^{-4}-15(\mathbb{D}V)V_{2y}^2V_yV^{-4}\\
&&+25(\mathbb{D}V)V_{2y}V_y^3V^{-5}-{29\over3}(\mathbb{D}V)V_y^5V^{-6}\Big\}.
\end{eqnarray*}

\subsection{Recursion operator of the SUSY HD equation}
In Section \ref{s3}, we have defined the reciprocal transformation between the SUSY HD equation \eqref{shd} and
the equation \eqref{spsckdv}, i.e.
\addtocounter{equation}{1}
\begin{align}
&\mathcal{D}=U^{-1/2}\mathbb{D}\tag{\theequation a}\label{eq41},\\
&\frac{\partial}{\partial t}=\frac{\partial}{\partial\tau}+\left(-\frac{1}{4}U_{2x}U+\frac{1}{8}U_x^2
+\frac{1}{4}(\mathcal{D}U_x)(\mathcal{D}U)\right)\frac{\partial}{\partial y}-\frac{1}{8}(\mathcal{D}U_{2x})U^{3/2}\mathbb{D}\tag{\theequation b}\label{eq42},\\
&S(y,\varrho,\tau)=U(x,\theta,t)\tag{\theequation c}\label{eq43}.
\end{align}
As in the classical case, above transformation is regarded as a B\"{a}cklund transformation involving independent variables
transformations. Let us first formulate the explicit independent variables changes of this transformations. It should be noted that the temporal variable is invariant, namely, $\tau=t$.

For a general superconformal transformation
\begin{equation*}
\mathcal{D}=G\mathbb{D},
\end{equation*}
we have
\begin{equation*}
\frac{\partial}{\partial x}=\left(\frac{\partial y}{\partial x}-\frac{\partial\varrho}{\partial x}\varrho\right)
\frac{\partial}{\partial y}+\frac{\partial\varrho}{\partial x}\mathbb{D}=G^2\frac{\partial}{\partial y}+(\mathcal{D}G)\mathbb{D},
\end{equation*}
which yield that
\addtocounter{equation}{1}
\begin{align}
\frac{\partial\varrho}{\partial x}&=(\mathcal{D}G), \tag{\theequation a}\label{eq44}\\
\frac{\partial y}{\partial x}&=G^2+\frac{\partial\varrho}{\partial x}\varrho. \tag{\theequation b}\label{eq45}
\end{align}
Integrating \eqref{eq44} and \eqref{eq45}, we have the explicit transformations
\begin{equation*}
\varrho=\int{\tt d}x\;(\mathcal{D}G)=(\mathcal{D}^{-1}G)\qquad\mbox{and}\qquad y=\int{\tt d}x\;\mathcal{D}[G(\mathcal{D}^{-1}G)]=\mathcal{D}^{-1}[G(\mathcal{D}^{-1}G)].
\end{equation*}

Hence, the reciprocal transformation \eqref{eq41}--\eqref{eq43} is equally reformulated as the following B\"{a}cklund transformation
\begin{equation}\label{eq46}
B(U,S)=U\big(x,\theta,t\big)-S\big(\mathcal{D}^{-1}[U^{-1/2}(\mathcal{D}^{-1}U^{-1/2})],(\mathcal{D}^{-1}U^{-1/2}),\tau=t\big)=0.
\end{equation}

To find a recursion operator for  the SUSY HD  equation \eqref{shd}, we first work with the equation \eqref{schw}. Noticing that
$V=S^{-1/2},\;t=\tau/4$ between \eqref{spsckdv} and \eqref{schw}, we may obtain for the equation \eqref{spsckdv} the following recursion operator
\begin{eqnarray*}
R(S)&=&\Big(-2S^{3/2}\Big)R(V)\mid_{V=S^{-1/2}}\Big(-\frac{1}{2}S^{-3/2}\Big)\\
&=&\Big(-2S^{3/2}\Big)P(V)\mid_{V=S^{-1/2}}J(V)\mid_{V=S^{-1/2}}\Big(-\frac{1}{2}S^{-3/2}\Big)\\
&=&S^{3/2}\mathbb{D}S^{1/2}\mathbb{D}S^{-1}\partial_y^{-1}S^{-1/2}\mathbb{D}S^{1/2}\mathbb{D}S^{1/2}\mathbb{D}S^{-1/2}
\partial_y^{-1}S^{-1}\mathbb{D}S^{1\over2}\mathbb{D}S^{1\over2}\mathbb{D}S^{-1}.
\end{eqnarray*}
Then the recursion operator of the SUSY HD equation \eqref{shd} is given by
\begin{eqnarray*}
R(U)&=&B_U^{-1}B_SR(S)\mid_{S=U,\;\mathbb{D}=U^{1/2}\mathcal{D}}B_S^{-1}B_U\\
&=&U^{3\over2}\mathcal{D}U^{1\over2}\mathcal{D}U^{-1}\partial_x^{-1}U\mathcal{D}U\mathcal{D}U^{-1}\mathcal{D}^{-1}U^{-1\over2}
\mathcal{D}^{-1}U^{-1\over2}\mathcal{D}U\mathcal{D}U\mathcal{D}U^{-1\over2}\\
&&\cdot\mathcal{D}^{-1}U^{-1\over2}\mathcal{D}^{-1}U^{-1}\mathcal{D}U\mathcal{D}U\mathcal{D}^3U\mathcal{D}^{-1}U^{-1/2}\mathcal{D}^{-1}U^{-3/2},
\end{eqnarray*}
where $B_U$ and $B_S$ are derivatives of $B(U,S)$ in directions $U$ and $S$ respectively, namely
\begin{equation*}
B_S=-1,\qquad B_U=U\partial_xU\mathcal{D}^{-1}U^{-1/2}\mathcal{D}^{-1}U^{-3/2}.
\end{equation*}
In Appendix \ref{app2}, a detailed calculation for $B_U$ is given.

The recursion operator $R(U)$ can be used to generate higher order flows from lower order ones. In fact, applying it to the 3rd order flow \eqref{shd} in the SUSY HD hierarchy, the 5th order flow is produced
\begin{eqnarray}\label{shd5}
U_{t_5}&=&{1\over4}U_{5x}U^5+{5\over4}U_{4x}U_xU^4+{5\over4}U_{3x}U_{2x}U^4+{5\over8}U_{3x}U_x^2U^3-{5\over8}(\mathcal{D}U_{4x})(\mathcal{D}U)U^4\nonumber\\
&&-{5\over8}(\mathcal{D}U_{3x})(\mathcal{D}U_{x})U^4-{35\over16}(\mathcal{D}U_{3x})(\mathcal{D}U)U_xU^3
-{15\over16}(\mathcal{D}U_{2x})(\mathcal{D}U_{x})U_xU^3\nonumber\\
&&-{15\over8}(\mathcal{D}U_{2x})(\mathcal{D}U)U_{2x}U^3
-{15\over32}(\mathcal{D}U_{2x})(\mathcal{D}U)U_x^2U^2-{15\over16}(\mathcal{D}U_{x})(\mathcal{D}U)U_{3x}U^3.
\end{eqnarray}

\subsection{Bi-Hamiltonian structure of the SUSY HD equation}
The Hamiltonian structures for SUSY HD equation \eqref{shd} is not known. The purpose of the subsection is to show that this system is actually a bi-Hamiltonian
system. As a first step, we derive a simple Hamiltonian structure within the framework of the $r$-matrix. To this end, we work with
 the super operator
\begin{equation*}
L=V\partial_x^2+\Phi\partial_x\mathcal{D}.
\end{equation*}
Applying the standard $r$-matrix approach to this super operator, one extract a simple Hamiltonian structure of the SUSY HD equation \eqref{shd} from the cubic Poisson tensor , which reads
\begin{equation*}
U_t= U^2{\cal D}^3U^2\frac{\delta }{\delta U}\int\verb"d"x\verb"d"\theta\; \left(-\frac{1}{8}U^{-1}U_x({\cal D}U)\right).
\end{equation*}
(for details, see Appendix \ref{app3}).

Let $P(U)=U^2{\cal D}^3U^2$, then the 5th order flow \eqref{shd5} is rewritten as a Hamiltonian system
\begin{equation*}
U_{t_5}=P(U)\frac{\delta}{\delta U}H(U)
\end{equation*}
with
\begin{eqnarray*}
H(U)&=&\int{\tt d}x{\tt d}\theta\;\Big(-\frac{1}{16}(\mathcal{D}U_x)U_x^2-\frac{1}{16}(\mathcal{D}U)U_{2x}U_x+\frac{1}{12}(\mathcal{D}U_{3x})U^2
+\frac{1}{8}(\mathcal{D}U_{2x})U_xU\\
&&+\frac{1}{8}(\mathcal{D}U_x)U_{2x}U+\frac{1}{24}(\mathcal{D}U)U_{3x}U+\frac{1}{32}(\mathcal{D}U)U_x^3U^{-1}\Big).
\end{eqnarray*}
Since
\begin{equation*}
R(U)U_t=U_{t_5}=P(U)\frac{\delta}{\delta U}H(U),
\end{equation*}
we have
\begin{equation*}
P^{-1}(U)R(U)U_t=\frac{\delta}{\delta U}H(U)
\end{equation*}
which indicates that $J(U)=P^{-1}(U)R(U)$ should be a candidate of a symplectic operator for the SUSY HD equation \eqref{shd}. However, in the present form even the validity of the skew-symmetric property is by no means clear, not to mention the sophisticated Jacobi identity. So we need to rewrite this operator.

First, we take the advantage of the computer algebra package SUSY2 \cite{susy2}. With its help, our operator $J(u)$ is simplified as
\begin{eqnarray*}
J(U)&=&U^{-2}\mathcal{D}-(\mathcal{D}U)U^{-3}+\frac{1}{4}U_xU^{-3/2}\Big(2\partial_x^{-1}U^{-3/2}-3\partial_x^{-1}(\mathcal{D}U)U^{-3/2}\Big)\\
&&+\frac{1}{8}U^{-3/2}\partial_x^{-1}\Big(-2U_x^2U^{-1}\partial_x^{-1}U^{-3/2}\mathcal{D}+3U_x^2U^{-1}\partial_x^{-1}(\mathcal{D}U)U^{-5/2}
-4U_xU^{-3/2}\mathcal{D}\\
&&+4(\mathcal{D}U_x)U^{-3/2}-2(\mathcal{D}U_x)(\mathcal{D}U)U^{-1}\partial_x^{-1}U^{-3/2}\mathcal{D}
+3(\mathcal{D}U_x)(\mathcal{D}U)U^{-1}\partial_x^{-1}(\mathcal{D}U)U^{-5/2}\\
&&+4(\mathcal{D}U)U_xU^{-5/2}-4(\mathcal{D}U)U^{-3/2}\partial_x^{-1}U_{2x}U^{-1}
+2(\mathcal{D}U)U^{-3/2}\partial_x^{-1}U_x^2U^{-2}\\
&&+4(\mathcal{D}U)U^{-3/2}\partial_x^{-1}(\mathcal{D}U_x)U^{-1}\mathcal{D}
+2(\mathcal{D}U)U^{-3/2}\partial_x^{-1}(\mathcal{D}U_x)(\mathcal{D}U)U^{-2}\\
&&-2(\mathcal{D}U)U^{-3/2}\partial_x^{-1}(\mathcal{D}U)U_xU^{-2}\mathcal{D}\Big)\\
&&+\frac{1}{4}(\mathcal{D}U_x)U^{-1}\partial_x^{-1}\Big(4U^{-2}-(\mathcal{D}U)U^{-3/2}\partial_x^{-1}U^{-3/2}\mathcal{D}+3(\mathcal{D}U)U^{-3/2}\partial_x^{-1}(\mathcal{D}U)U^{-5/2}\Big)\\
&&-\frac{1}{8}(\mathcal{D}U)U_xU^{-2}\partial_x^{-1}\Big(4U^{-2}-(\mathcal{D}U)U^{-3/2}\partial_x^{-1}U^{-3/2}\mathcal{D}+3(\mathcal{D}U)U^{-3/2}\partial_x^{-1}(\mathcal{D}U)U^{-5/2}\Big).
\end{eqnarray*}
By recombining terms properly, the operator $J(U)$ can be rewritten neatly in the product form
\begin{equation}
J(U)=U^{-3/2}\mathcal{D}^{-1}U^{-1/2}
\mathcal{D}^{-1}U\mathcal{D}^{5}U\mathcal{D}^{-1}U^{-1/2}\mathcal{D}^{-1}U^{-3/2}.
\end{equation}

It is now easy to see that the operator $J(U)$ is skew-symmetric, so to confirm that $J(U)$ is symplectic, it is sufficient to prove the Jacobi identity for it. Taking account of the form of $J(U)$, it is easier to verify  the Jacobi identity for $J^{-1}(U)$.
\begin{prop}\label{prop2}
The operator
\begin{equation*}
Q\equiv J^{-1}(U)=U^{3/2}\mathcal{D}U^{1/2}\mathcal{D}U^{-1}\mathcal{D}^{-5}U^{-1}\mathcal{D}U^{1/2}\mathcal{D}U^{3/2}
\end{equation*}
satisfies the Jacobi identity
\begin{equation}\label{jac}
\langle\alpha,Q^{\prime}_{Q(\beta)}(\gamma)\rangle+\langle\beta,Q^{\prime}_{Q(\gamma)}(\alpha)\rangle
+\langle\gamma,Q^{\prime}_{Q(\alpha)}(\beta)\rangle=0,
\end{equation}
where $\alpha$, $\beta$ and $\gamma$ are fermionic testing functions. Hence $J^{-1}(U)$ does qualify as a Hamiltonian operator of
the SUSY HD equation \eqref{shd}.
\end{prop}

The proof of above result is tricky and long, so we postpone it to Appendix \ref{app4}.

\section{SUSY reciprocal transformation: SUSY Kawamoto case}\label{s5}
In this section, we consider fifth order equations. The analog of the HD equation now is the Kawamoto equation \cite{kawa}
\begin{equation}\label{kwmt}
v_t=v^5v_{5x}+5v^4v_xv_{4x}+\frac{5}{2}v^4v_{2x}v_{3x}+\frac{15}{4}v^3v_x^2v_{3x},
\end{equation}
which is integrable. Indeed, its Lax operator  reads as \cite{das}
\begin{equation*}
L=v^3\partial_x^3+3v_xv^2\partial_x^2~.
\end{equation*}

As Kawamoto showed, the equation (\ref{kwmt}) is reciprocally associated to the following equation
\begin{equation}\label{mskk}
w_t=w_{5x}-5w_xw_{3x}-5w_{2x}^2-5w_x^3-20ww_xw_{2x}-5w^2w_{3x}+5w^4w_x~.
\end{equation}

It is a well known fact that \cite{kawa, fordy}, through the Miura transformations
\[
u=w_x-w^2,
\]
and
\[
u=-w_x-\frac{1}{2}w^2
\]
respectively, the equation (\ref{mskk}), sometimes known as
Fordy-Gibbons equation,  is the common modification, of the
Sawada-Kotera equation
\begin{equation*}
u_t=u_{5x}+5u_{3x}u+5u_{2x}u_x+5u_xu^2
\end{equation*}
and of the Kaup-Kupershmidt equation
\begin{equation}\label{kk}
u_t=u_{5x}+10u_{3x}u+25u_{2x}u_x+20u_xu^2.
\end{equation}

Now let us turn to the supersymmetric case. The equation (\ref{kwmt}) is embedded in its supersymmetric analogy as
\begin{eqnarray}
V_t&=&V^5V_{5x}+5V^4V_{4x}V_x+\frac{5}{2}V^4V_{3x}V_{2x}+\frac{15}{4}V^3V_{3x}V_x^2
-\frac{5}{2}V^4(\mathcal{D}V_{4x})(\mathcal{D}V)\nonumber\\
&&-5V^4(\mathcal{D}V_{3x})(\mathcal{D}V_x)-\frac{15}{2}V^3(\mathcal{D}V_{3x})(\mathcal{D}V)V_x
-\frac{15}{2}V^3(\mathcal{D}V_{2x})(\mathcal{D}V_x)V_x\nonumber\\
&&-\frac{15}{4}V^3(\mathcal{D}V_{2x})(\mathcal{D}V)V_{2x}
-\frac{15}{8}V^2(\mathcal{D}V_{2x})(\mathcal{D}V)V_x^2\label{skwmt}~,
\end{eqnarray}
whose integrability follows from the Lax representation
\begin{equation*}
\left[\frac{\partial}{\partial t}+9(L_k^{5/3})_{\geq 3}, L_k\right]=0,
\end{equation*}
where the Lax operator is given by
\begin{equation}\label{skwmtlax}
L_k=V^{3/2}\mathcal{D}^3V^{3/2}\mathcal{D}^3
\end{equation}
and $V$ is a bosonic super field.

The equation (\ref{skwmt}) is our  supersymmetric Kawamoto equation
and the rest of this section will be devoted to its relationship
with the supersymmetric Sawada-Kotera equation. Let us introduce
the super Liouville transformation
\begin{equation*}
\mathcal{D}=S^{-1/2}\mathbb{D},\qquad S(y,\varrho,\tau)=V(x,\theta,t),
\end{equation*}
then, the operator $L_k$ takes the form
\begin{eqnarray}
L_{msk}&=&\partial_y^3-3(\mathbb{D}\Psi)\partial_y^2+[2(\mathbb{D}\Psi)^2-2(\mathbb{D}\Psi_y)-\Psi_y\Psi]\partial_y\nonumber\\
&&+[\Psi_y(\mathbb{D}\Psi)-\Psi_{2y}+\Psi(\mathbb{D}\Psi_y)]\mathbb{D},\label{smsklax}
\end{eqnarray}
where the fermionic super field $\Psi$ is related with the bosonic $W$ by the super Cole-Hopf transformation
\begin{equation}\label{scole2}
\Psi=-\mathbb{D}\log(S^{-1/2})~.
\end{equation}

Using the transformed operator $L_{msk}$ we consider the following Lax equation
\begin{equation*}
\left[\frac{\partial}{\partial \tau}+9(L_{msk}^{5/3})_{\geq 1},L_{msk}\right]=0~,
\end{equation*}
which provides us the SUSY nonlinear evolution equation
\begin{eqnarray}\label{smsk}
\Psi_{\tau} &=& {\mathbb{ D}} \big [ ({\mathbb{ D}}  \Psi_{4y})-5({\mathbb{D}}\Psi_{2y})({\mathbb{D}}\Psi_y)-
5({\mathbb{D}}\Psi_{2y})({\mathbb{D}}\Psi)^2-5({\mathbb{D}}\Psi_y)^2({\mathbb{D}}\Psi)+ ({\mathbb{D}}\Psi)^5\nonumber\\
&&-5\Psi_{3y}\Psi_y -5\Psi_{2y}\Psi_y({\mathbb{D}}\Psi)-10\Psi_y\Psi({\mathbb{D}}\Psi_{2y}) -10\Psi_y \Psi
({\mathbb{D}}\Psi_y)({\mathbb{D}}\Psi)\big ].
\end{eqnarray}

The above equation is a modification of the SUSY SK equation proposed in Ref.\;\onlinecite{tian}. To see it, we can further bring  $L_{msk}$, through  gauge transformation, to a new operator, i.e.
\begin{equation}\label{gaugeT}
 L_{sk}
 = e^{-(\mathbb{D}^{-1}\Psi)} L_{msk}\; e^{(\mathbb{D}^{-1}\Psi)} = \Big ({\mathbb{D}}^3 + \Psi_y - \Psi({\mathbb{D}}\Psi) \Big )^2= \big ({\cal D}^3 + \Phi\big)^2,
\end{equation}
which is nothing but the Lax operator for the SUSY SK equation which reads as
\begin{equation}\label{ssk}
\Phi_\tau=\Phi_{5y}+5\Phi_{3y}(\mathbb{D}\Phi)+5\Phi_{2y}(\mathbb{D}\Phi_y)+5\Phi_y(\mathbb{D}\Phi)^2.
\end{equation}
The Lax operator and infinite number of supersymmetric conserved densities for this equation have been found in Ref.\;\onlinecite{tian} while  the odd bi-Hamiltonian structure was constructed in Ref.\;\onlinecite{pop}.
Due to the existence of the Miura transformation, which follow from the equation (\ref{gaugeT})
\begin{equation}\label{skmiu}
 \Phi=\Psi_y - \Psi({\mathbb D}\Psi).
\end{equation}
based on the odd bi-Hamiltonian structure of the SUSY SK equation,
 we will calculate the odd bi-Hamiltonian  structure of the modified equation (\ref{smsk}) in next section.

To complete the construction of the reciprocal transformation, one
has to find the relation between $\partial_t$ and
$\partial_\tau$. As in the SUSY HD case, this can
be done in two ways, either using linear problem or making use of
conservation law. We now take the latter approach and will show that
the result follows from Proposition 1. To do it, we notice that the
equation (\ref{skwmt}) admits the conservation law
\begin{eqnarray*}
\frac{\partial}{\partial
t}\left(V^{-1/2}\right)&=&\frac{1}{8}\mathcal{D}\Big(
-4(\mathcal{D}V_{4x})V^{7/2}-16(\mathcal{D}V_{3x})V_xV^{5/2}-14(\mathcal{D}V_{2x})V_{2x}V^{5/2}\\
&&-5(\mathcal{D}V_{2x})V_x^2V^{3/2}+20(\mathcal{D}V_{2x})(\mathcal{D}V_x)(\mathcal{D}V)V^{3/2}-4(\mathcal{D}V)V_{4x}V^{5/2}\\
&&+4(\mathcal{D}V_x)V_{3x}V^{5/2}-10(\mathcal{D}V)V_{3x}V_xV^{3/2}\Big),
\end{eqnarray*}
which implies that we can identify
\[
G=V^{-1/2},
\]
and
\begin{eqnarray*}
\Xi&=&\frac{1}{8}\Big(-4(\mathcal{D}V_{4x})V^{7/2}-16(\mathcal{D}V_{3x})V_xV^{5/2}-14(\mathcal{D}V_{2x})V_{2x}V^{5/2}\\
&&-5(\mathcal{D}V_{2x})V_x^2V^{3/2}+20(\mathcal{D}V_{2x})(\mathcal{D}V_x)(\mathcal{D}V)V^{3/2}-4(\mathcal{D}V)V_{4x}V^{5/2}\\
&&+4(\mathcal{D}V_x)V_{3x}V^{5/2}-10(\mathcal{D}V)V_{3x}V_xV^{3/2}\Big).
\end{eqnarray*}
Integrating $2\Xi V^{-1/2}$ yields
\begin{eqnarray*}
W&=&\frac{1}{4}\Big(-4V_{4x}V^3-8V_{3x}V_xV^2+8(\mathcal{D}V_{3x})(\mathcal{D}V)V^2
+V_{2x}V_x^2V\\
&&-V_{2x}^2V^2-\frac{1}{4}V_x^4+12(\mathcal{D}V_{2x})(\mathcal{D}V_x)V^2+6(\mathcal{D}V_{2x})(\mathcal{D}V)V_xV\\
&&+2(\mathcal{D}V_x)(\mathcal{D}V)V_{2x}V-(\mathcal{D}V_x)(\mathcal{D}V)V_x^2\Big).
\end{eqnarray*}
Therefore, our reciprocal transformation in this case reads as
\addtocounter{equation}{1}
\begin{align}
&\mathcal{D}=V^{-1/2}\mathbb{D} \tag{\theequation a},\label{sucon}\\
&\frac{\partial}{\partial t}=W\frac{\partial}{\partial y}+\Xi\mathbb{D}+\frac{\partial}{\partial\tau}.\tag{\theequation b}\label{tpartkw}
\end{align}

Under the transformation (\ref{sucon}) and (\ref{tpartkw}), the
equation (\ref{skwmt}) is converted into
\begin{eqnarray}
S_\tau&=&S_{5y}-5S_{4y}S_yS^{-1}-{25\over2}S_{3y}S_{2y}S^{-1}+{85\over4}S_{3y}S_y^2S^{-2}+{145\over4}S_{2y}^2S_yS^{-2}\nonumber\\
&&-{265\over4}S_{2y}S_y^3S^{-3}+{405\over16}S_y^5S^{-4}-{5\over2}(\mathbb{D}S_{3y})(\mathbb{D}S_y)S^{-1}\nonumber\\
&&+{5\over2}(\mathbb{D}S_{3y})(\mathbb{D}S)S_yS^{-2}+{25\over4}(\mathbb{D}S_{2y})(\mathbb{D}S_{y})S_yS^{-2}\nonumber\\
&&-{25\over4}(\mathbb{D}S_{2y})(\mathbb{D}S)S_y^2S^{-3}-5(\mathbb{D}S_{y})(\mathbb{D}S)S_{3y}S^{-2}\nonumber\\
&&+{25\over2}(\mathbb{D}S_{y})(\mathbb{D}S)S_{2y}S_yS^{-3}-{15\over4}(\mathbb{D}S_{y})(\mathbb{D}S)S_y^3S^{-4},\label{scmsk}
\end{eqnarray}
which is related to the equation (\ref{smsk}) through the transformation (\ref{scole2}).

\section{Recursion operator of the SUSY Kawamoto equation}\label{s6}
In last section, a SUSY Kawamoto equation \eqref{skwmt} was proposed, and was shown to be transformed into the equation \eqref{scmsk}
under the reciprocal transformation \eqref{sucon}\eqref{tpartkw}. This reciprocal transformation supplies us a B\"{a}cklund transformation
between these two equations, formulated as
\begin{equation}\label{kwmtbt}
B(V,S)=V\big(x,\theta,t\big)-S\big(\mathcal{D}^{-1}[V^{-1/2}(\mathcal{D}^{-1}V^{-1/2})],(\mathcal{D}^{-1}V^{-1/2}),\tau=t\big)=0.
\end{equation}
Through this B\"{a}cklund transformation, the recursion operator, and even odd Hamiltonian structure of the SUSY Kawamoto equation can be
constructed as we did in the case of the SUSY HD equation.

We should start with the SUSY SK equation \eqref{ssk}, whose odd bi-Hamiltonian structure reads \cite{pop}
\begin{eqnarray*}
\Omega(\Phi)\Phi_\tau=\frac{\delta}{\delta\Phi}H_1(\Phi),\qquad
\Phi_\tau=\Lambda(\Phi)\frac{\delta}{\delta\Phi}H_0(\Phi),
\end{eqnarray*}
where the symplectic operator $\Omega(\Phi)$, the odd Hamiltonian operator $\Lambda(\Phi)$ and the corresponding Hamiltonians are
\begin{eqnarray*}
\Omega(\Phi)&=&\partial_y^2+(\mathbb{D}\Phi)-\partial_y^{-1}(\mathbb{D}\Phi_y)+\partial_y^{-1}\Phi_y\mathbb{D}+\Phi_y\mathbb{D}^{-1}
=\mathbb{D}^{-1}\Big(\mathbb{D}^3+\Phi\Big)\Big(\mathbb{D}^3+\Phi\Big)\mathbb{D}^{-1},\\
\Lambda(\Phi)&=&\Big(\partial_y^2\mathbb{D}+2\partial_y\Phi+2\Phi\partial_y+\mathbb{D}\Phi\mathbb{D}\Big)\partial_y^{-1}
\Big(\partial_y^2\mathbb{D}+2\partial_y\Phi+2\Phi\partial_y+\mathbb{D}\Phi\mathbb{D}\Big),\\
H_0(\Phi)&=&\int\verb"d"y\verb"d"\varrho\left(-\frac{1}{2}\Phi_y\Phi\right),\\
H_1(\Phi)&=&\int\verb"d"y\verb"d"\varrho\Big(-\frac{1}{2}\Phi_{7y}\Phi-4\Phi_{3y}\Phi(\mathbb{D}\Phi_{2y})\\
&&\qquad\qquad\qquad-\Phi_y\Phi\left[2(\mathbb{D}\Phi_{4y})
+10(\mathbb{D}\Phi_{2y})(\mathbb{D}\Phi)+5(\mathbb{D}\Phi_y)^2+\frac{4}{3}(\mathbb{D}\Phi)^3\right]\Big).
\end{eqnarray*}

The odd bi-Hamiltonian structure of the SUSY MSK equation \eqref{smsk} is constructed through the Miura transformation \eqref{skmiu}.
We skip details and present the result directly,
\begin{equation*}
\Omega(\Psi)\Psi_\tau=\frac{\delta}{\delta\Psi}H_1(\Psi),\qquad
\Psi_\tau=\Lambda(\Psi)\frac{\delta}{\delta\Psi}H_0(\Psi),
\end{equation*}
where
\begin{eqnarray*}
\Omega(\Psi)&=&-(\partial_y+2\mathbb{D}\Psi+\Psi\mathbb{D})\mathbb{D}^{-1}(\mathbb{D}-\Psi)\mathbb{D}(\mathbb{D}+\Psi)
(\mathbb{D}-\Psi)\mathbb{D}(\mathbb{D}+\Psi)\mathbb{D}^{-1}(\partial_y-2\Psi\mathbb{D}-\mathbb{D}\Psi),\\
\Lambda(\Psi)&=&-\mathbb{D}(\partial_y+2\mathbb{D}\Psi+\Psi\mathbb{D})\partial_y^{-1}(\partial_y-2\Psi\mathbb{D}-\mathbb{D}\Psi)\mathbb{D},
\end{eqnarray*}
and
\begin{eqnarray*}
H_0(\Psi)=H_0(\Phi)\mid_{\Phi=\Psi_y-\Psi(\mathcal{D}\Psi)}\qquad H_1(\Psi)=H_1(\Phi)\mid_{\Phi=\Psi_y-\Psi(\mathcal{D}\Psi)}.
\end{eqnarray*}

Taking account of the equation \eqref{scmsk}, it is possible to formulate its odd bi-Hamiltonian structure from that of the SUSY MSK equation through the transformation \eqref{scole2}. However, we are only interested in its recursion operator here, which is given by
\begin{eqnarray*}
R(S)&=&\Psi_S^{-1}\cdot\Lambda(\Psi)\mid_{\Psi=-\mathbb{D}\log S^{-1/2}}\Omega(\Psi)\mid_{\Psi=-\mathbb{D}\log S^{-1/2}}\cdot\Psi_S\\
&=&S^{1/2}\mathbb{D}S^{-1/2}\mathbb{D}S\partial_y^{-1}S\mathbb{D}S^{-1/2}\mathbb{D}S^{-1/2}\mathbb{D}S^{-1/2}
\mathbb{D}S^{-1/2}\mathbb{D}S\mathbb{D}^{-1}S^{1/2}\\
&&\cdot\mathbb{D}S^{-1/2}\mathbb{D}S^{-1/2}\mathbb{D}S\mathbb{D}S^{-1/2}\mathbb{D}S^{-1/2}\mathbb{D}S^{1/2}
\mathbb{D}^{-1}S\mathbb{D}S^{-1/2}\mathbb{D}S^{-1/2}\mathbb{D}S^{-1}.
\end{eqnarray*}
From it, the recursion operator of the SUSY Kawamoto equation is obtained
\begin{eqnarray*}
R(V)&=&B_V^{-1}B_SR(S)\mid_{S=V,\;\mathbb{D}=V^{1/2}\mathcal{D}}B_S^{-1}B_V\\
&=&V^{5/2}\mathcal{D}^5V\mathcal{D}^{-1}V^{1/2}\mathcal{D}^3V^{3/2}\mathcal{D}^3V^{1/2}
\mathcal{D}^{-1}V\mathcal{D}^5V\mathcal{D}^{-1}V^{-1/2}\mathcal{D}^{-1}V^{-3/2}.
\end{eqnarray*}

A simple, but interesting observation is that the recursion operator admits the decomposition
\begin{equation*}
R(V)=\Lambda(V)\cdot\Pi^{-1}(V),
\end{equation*}
where
\begin{eqnarray*}
\Pi(V)&=&V^{3/2}\mathcal{D}V^{1/2}\mathcal{D}V^{3/2},\\
\Lambda(V)&=&V^{5/2}\mathcal{D}^5V\mathcal{D}^{-1}V^{1/2}\mathcal{D}^3V^{3/2}\mathcal{D}^3V^{1/2}
\mathcal{D}^{-1}V\mathcal{D}^5V^{5/2}.
\end{eqnarray*}
Also,  the SUSY Kawamoto equation \eqref{skwmt} is rewritten as
\begin{equation}\label{bi_susykawa}
V_t=\Pi(V)\frac{\delta}{\delta V}H_1(V)=\Lambda(V)\frac{\delta}{\delta V}H_0(V),
\end{equation}
where
\begin{eqnarray*}
H_0(V)&=&\int\verb"d"x\verb"d"\theta\;\Big(-2V^{-1/2}\Big),\\
H_1(V)&=&\int\verb"d"x\verb"d"\theta\;\Big(\frac{2}{7}V_{4x}V^{5\over2}+{4\over7}V_{3x}V_xV^{3\over2}+{1\over14}V_{2x}^2V^{3\over2}
-{1\over14}V_{2x}V_x^2V^{1\over2}+{1\over56}V_x^4V^{-1\over2}\\
&&-{4\over7}(\mathcal{D}V_{3x})(\mathcal{D}V)V^{3\over2}
-{6\over7}(\mathcal{D}V_{2x})(\mathcal{D}V_{x})V^{3\over2}-{3\over7}(\mathcal{D}V_{2x})(\mathcal{D}V)V_xV^{1\over2}\\
&&-{1\over7}(\mathcal{D}V_{x})(\mathcal{D}V)V_{2x}V^{1\over2}+{1\over14}(\mathcal{D}V_{x})(\mathcal{D}V)V_x^2V^{-1\over2}\Big).
\end{eqnarray*}

The skew-symmetry of the operator $\Pi(V)$ is evident. So to show that this operator is an odd Hamiltonian operator, we only need to verify the Jacobi identity for it, this is contained in the next Proposition.

\begin{prop}
The operator $\Pi(V)$ satisfies the Jacobi identity
\begin{equation*}
\langle A,\; \Pi(V)'_{\Pi(V)B}(C)\rangle+\langle B,\; \Pi(V)'_{\Pi(V)C}(A)\rangle+\langle C,\; \Pi(V)'_{\Pi(V)A}(B)\rangle=0
\end{equation*}
where $A$, $B$ $C$ are bosonic testing functions.
Hence, it does qualify as an odd Hamiltonian operator for the SUSY Kawamoto equation (\ref{skwmt}).
\end{prop}

\noindent{\bf Proof:} Since
\begin{eqnarray*}
&&\langle A,\; \Pi(V)'_{\Pi(V)B}(C)\rangle\\
&=&\int\verb"d"x\verb"d"\theta\Big\{\frac{3}{2}AV^{1/2}[\Pi(V)B]\mathcal{D}V^{1/2}\mathcal{D}V^{3/2}C
+\frac{1}{2}AV^{3/2}\mathcal{D}V^{-1/2}[\Pi(V)B]\mathcal{D}V^{3/2}C\\
&&~~~~~~~~~~~+\frac{3}{2}AV^{3/2}\mathcal{D}V^{1/2}\mathcal{D}V^{1/2}[\Pi(V)B]C\Big\}\\
&=&\int\verb"d"x\verb"d"\theta\Big\{\frac{3}{2}V^{-1}A[\Pi(V)B][\Pi(V)C]-\frac{3}{2}[\Pi(V)A][\Pi(V)B]C\\
&&~~~~~~~~~~~-\frac{1}{2}[\mathcal{D}V^{1/2}\mathcal{D}V^{3/2}B][V^{1/2}\mathcal{D}V^{3/2}A][V^{1/2}\mathcal{D}V^{3/2}C]\Big\},
\end{eqnarray*}
we have
\begin{eqnarray*}
&&\langle A,\; \Pi(V)'_{\Pi(V)B}C\rangle+\langle B,\;\Pi(V)'_{\Pi(V)C}A\rangle+\langle C,\; \Pi(V)'_{\Pi(V)A}B\rangle\\
&=&\int\verb"d"x\verb"d"\theta\Big\{\frac{3}{2}V^{-1}A[\Pi(V)B][\Pi(V)C]-\frac{3}{2}[\Pi(V)A][\Pi(V)B]C+\frac{3}{2}V^{-1}B[\Pi(V)C][\Pi(V)A]\\
&&~~~~~~~~~~~-\frac{3}{2}[\Pi(V)B][\Pi(V)C]A+\frac{3}{2}V^{-1}C[\Pi(V)A][\Pi(V)B]-\frac{3}{2}[\Pi(V)C][\Pi(V)A]B\\
&&~~~~~~~~~~~-\frac{1}{2}[\mathcal{D}V^{1/2}\mathcal{D}V^{3/2}B][V^{1/2}\mathcal{D}V^{3/2}A][V^{1/2}\mathcal{D}V^{3/2}C]\\
&&~~~~~~~~~~~-\frac{1}{2}[\mathcal{D}V^{1/2}\mathcal{D}V^{3/2}C][V^{1/2}\mathcal{D}V^{3/2}B][V^{1/2}\mathcal{D}V^{3/2}A]\\
&&~~~~~~~~~~~-\frac{1}{2}[\mathcal{D}V^{1/2}\mathcal{D}V^{3/2}A][V^{1/2}\mathcal{D}V^{3/2}C][V^{1/2}\mathcal{D}V^{3/2}B]\Big\}\\
&=&\int\verb"d"x\verb"d"\theta\;\mathcal{D}\left\{\frac{1}{2}[V^{1/2}\mathcal{D}V^{3/2}A][V^{1/2}\mathcal{D}V^{3/2}B][V^{1/2}\mathcal{D}V^{3/2}C]\right\}\\
&=&0
\end{eqnarray*}
The Jacobi identity holds.  \hfill $\Box$

{\bf Remark}: The Jacobi identity in the present case also  could be checked directly with the computer algebra package SUSY2.
\bigskip

It is natural to consider \eqref{bi_susykawa} as an odd bi-Hamiltonian system, but we have not yet checked the Jacobi identity
for the operator $\Lambda(V)$.

\section{Conclusion}
The reciprocal link, sometimes  also named as hodograph
transformation,  is a useful instrument which allows us to
transform one equation  to another equation which in some cases is
a  well known equation. One would like to say the same in the
supersymmetric case but then the situation is more complicated. In
this paper we constructed the supersymmetric analogy of the
reciprocal link between the  supersymmetric Harry Dym and MKdV equations.
We constructed a Lax representation for supersymmetric Kawamoto
equation and established the supersymmetric reciprocal link to the
supersymmetric Sawada-Kotera equation. Exploiting these links we found the recursion
operator as well as the (odd) bi-Hamiltonian structure for the supersymmetric Harry Dym (Kawamoto)
equation. This opens us the door to further investigations of the higher or negative flows
for the supersymmetric Harry Dym equations.

\section*{Acknowledgement}

We should like to thank Decio Levi for many interesting discussions. This
work is supported by National Natural Science Foundation of China
with grant numbers 10731080 and 10971222.

\appendix
\section{}\label{app1}
In addition to the conservation law (\ref{conser}), the SUSY HD equation (\ref{shd}) has another one given by
\begin{equation}
\frac{\partial}{\partial t}\Big(U^{-1}\Big)=\mathcal{D}\left(\frac{1}{8}(\mathcal{D}U)U_{2x}
-\frac{1}{8}(\mathcal{D}U_x)U_x-\frac{1}{4}(\mathcal{D}U_{2x})U\right).
\end{equation}
Additionally, a potential can be introduced for the quantity
\begin{equation*}
2\left(\frac{1}{8}(\mathcal{D}U)U_{2x}
-\frac{1}{8}(\mathcal{D}U_x)U_x-\frac{1}{4}(\mathcal{D}U_{2x})U\right)U^{-1}
\end{equation*}
i.e.
\begin{eqnarray}
\mathcal{D}\left(\frac{1}{4}(\mathcal{D}U_x)(\mathcal{D}U)U^{-1}-\frac{1}{2}U_{2x}\right)
=2\left(\frac{1}{8}(\mathcal{D}U)U_{2x}
-\frac{1}{8}(\mathcal{D}U_x)U_x-\frac{1}{4}(\mathcal{D}U_{2x})U\right)U^{-1}.
\end{eqnarray}

According to the Proposition \ref{prop1}, one can apply the reciprocal transformation
\begin{eqnarray}
\mathcal{D}&=&U^{-1}\mathbb{D},\label{rti1}\\
\frac{\partial}{\partial t}&=&\left(\frac{1}{4}(\mathcal{D}U_x)(\mathcal{D}U)U^{-1}-\frac{1}{2}U_{2x}\right)\frac{\partial}{\partial y}\nonumber\\
&&+\left(\frac{1}{8}(\mathcal{D}U)U_{2x}
-\frac{1}{8}(\mathcal{D}U_x)U_x-\frac{1}{4}(\mathcal{D}U_{2x})U\right)\mathbb{D}+\frac{\partial}{\partial \tau}\label{rti2}
\end{eqnarray}
to the SUSY HD equation (\ref{shd}).

A direct calculation gives us
\begin{eqnarray}
U_\tau&=&\frac{1}{4}U_{3y}U^{-3}-\frac{3}{2}U_{2y}U_yU^{-4}+\frac{3}{8}(\mathbb{D}U_{2y})(\mathbb{D}U)U^{-4}\nonumber\\
&&+\frac{3}{2}U_y^3U^{-5}-\frac{3}{2}(\mathbb{D}U_y)(\mathbb{D}U)U_yU^{-5}
\end{eqnarray}
which, by $U=\hat{U}^{-1}$, is transformed to the SUSY HD equation
\begin{equation*}
\hat{U}_\tau=\frac{1}{4}\hat{U}_{3y}\hat{U}^3-\frac{3}{8}(\mathbb{D}\hat{U}_{2y})(\mathbb{D}\hat{U})U^2.
\end{equation*}

The invariance of the SUSY HD equation (\ref{shd}) under the SUSY reciprocal transformation (\ref{rti1}) and (\ref{rti2}) can be viewed as a
SUSY generalization of the invariance of the HD equation (\ref{hd}) under the transformation
\begin{equation*}
\verb"d"y=u^{-2}\verb"d"x+2u_{2x}\verb"d"t,\qquad \verb"d"\tau=\verb"d"t
\end{equation*}
which follows from the conservation law
\begin{equation*}
\frac{\partial}{\partial t}\Big(u^{-2}\Big)=\frac{\partial}{\partial x}\Big(2u_{2x}\Big),
\end{equation*}
and was first reported in Ref.\;\onlinecite{rogers}.

\section{}\label{app2}
The Fr\'{e}chet derivative of the B\"{a}cklund transformation \eqref{eq46} in the direction $U$ is define by
\begin{eqnarray*}
B_U[Q]&=&\frac{\tt{d}}{\tt{d}\epsilon}\mid_{\epsilon=0}B(U+\epsilon Q,S)\\
&=&\frac{\tt{d}}{\tt{d}\epsilon}\mid_{\epsilon=0}\Big\{U+\epsilon Q-S\big(\mathcal{D}^{-1}[(U+\epsilon Q)^{-1/2}\mathcal{D}^{-1}(U+\epsilon Q)^{-1/2}],\mathcal{D}^{-1}(U+\epsilon Q)^{-1/2}\big)\Big\}.
\end{eqnarray*}
Direct calculation leads to
\begin{eqnarray*}
B_U[Q]&=&Q-\left\{-\frac{1}{2}\mathcal{D}^{-1}\Big[U^{-3/2}Q(\mathcal{D}^{-1}U^{-1/2})+U^{-1/2}(\mathcal{D}^{-1}U^{-3/2}Q)\Big]\right\}S_y\\
&&\qquad-\left\{-\frac{1}{2}(\mathcal{D}^{-1}U^{-3/2}Q)\right\}S_\varrho\\
&=&Q+\frac{1}{2}\mathcal{D}^{-1}\Big[U^{-3/2}Q(\mathcal{D}^{-1}U^{-1/2})+U^{-1/2}(\mathcal{D}^{-1}U^{-3/2}Q)\Big]S_y\\
&&\qquad+\frac{1}{2}(\mathcal{D}^{-1}U^{-3/2}Q)\big[(\mathbb{D}S)-\varrho S_y\big]\\
&=&Q+\frac{1}{2}\mathcal{D}^{-1}\Big[U^{-3/2}Q(\mathcal{D}^{-1}U^{-1/2})+U^{-1/2}(\mathcal{D}^{-1}U^{-3/2}Q)\Big]S_y\\
&&\qquad+\frac{1}{2}(\mathcal{D}^{-1}U^{-3/2}Q)(\mathbb{D}\hat{U})+\frac{1}{2}(\mathcal{D}^{-1}U^{-1/2})(\mathcal{D}^{-1}U^{-3/2}Q)S_y\\
&=&Q+\frac{1}{2}\mathcal{D}^{-1}\Big[U^{-3/2}Q(\mathcal{D}^{-1}U^{-1/2})+U^{-1/2}(\mathcal{D}^{-1}U^{-3/2}Q)\Big]UU_x\\
&&\qquad+\frac{1}{2}(\mathcal{D}^{-1}U^{-3/2}Q)(\mathcal{D}U)U^{1/2}+\frac{1}{2}(\mathcal{D}^{-1}U^{-1/2})(\mathcal{D}^{-1}U^{-3/2}Q)UU_x.
\end{eqnarray*}
Since
\begin{equation*}
(\mathcal{D}^{-1}U^{-1/2})(\mathcal{D}^{-1}U^{-3/2}Q)+\mathcal{D}^{-1}\Big[(\mathcal{D}^{-1}U^{-1/2})U^{-3/2}Q\Big]
=\Big[\mathcal{D}^{-1}U^{-1/2}(\mathcal{D}^{-1}U^{-3/2}Q)\Big],
\end{equation*}
$B_U(Q)$ is simplified as
\begin{equation*}
B_U[Q]=Q+UU_x\Big[\mathcal{D}^{-1}U^{-1/2}(\mathcal{D}^{-1}U^{-3/2}Q)\Big]-\frac{1}{2}U^{1/2}(\mathcal{D}U)(\mathcal{D}^{-1}U^{-3/2}Q).
\end{equation*}
Replacing $(\mathcal{D}U)$ by $(\mathcal{D}\cdot U-U\cdot\mathcal{D})$ and $U_x$ by $(\partial_x\cdot U-U\cdot\partial_x)$ respectively, then
\begin{eqnarray*}
B_U[Q]&=&\frac{3}{2}Q+(U\partial_xU\mathcal{D}^{-1}U^{-1/2}\mathcal{D}^{-1}U^{-3/2}Q)
-\left[U^2\mathcal{D}U^{-1/2}+\frac{1}{2}U^{1/2}\mathcal{D}U\right](\mathcal{D}^{-1}U^{-3/2}Q)\\
&=&\frac{3}{2}Q+(U\partial_xU\mathcal{D}^{-1}U^{-1/2}\mathcal{D}^{-1}U^{-3/2}Q)\\
&&\qquad-\left[-\frac{1}{2}U^{1/2}(\mathcal{D}U)+U^{3/2}\mathcal{D}+\frac{1}{2}U^{3/2}\mathcal{D}
+\frac{1}{2}U^{1/2}(\mathcal{D}U)\right](\mathcal{D}^{-1}U^{-3/2}Q)\\
&=&\frac{3}{2}Q+(U\partial_xU\mathcal{D}^{-1}U^{-1/2}\mathcal{D}^{-1}U^{-3/2}Q)-\frac{3}{2}Q\\
&=&(U\partial_xU\mathcal{D}^{-1}U^{-1/2}\mathcal{D}^{-1}U^{-3/2}Q).
\end{eqnarray*}

\section{}\label{app3}
The algebra $\mathfrak{g}$ of super pseudodifferential operators admits three kinds of subalgebra decompositions
\begin{eqnarray*}
\mathfrak{g}=\mathfrak{g}_{\geq k}\oplus \mathfrak{g}_{< k}
\end{eqnarray*}
where $k=0, 1, 3$. We only consider the case: $k=3$. It was pointed out \cite{kono} that
\begin{eqnarray}
\mathscr{P}_3(L)\nabla H&=&\quad 2[P_{\geq 3}(L\nabla HL),L]-2LP^*_{<3}([\nabla H,L])L\label{cu1}\\
&=&-2[P_{< 3}(L\nabla HL),L]+2LP^*_{\geq 3}([\nabla H,L])L.\label{cu2}
\end{eqnarray}
defines a Poisson tensor on $\mathfrak{g}$.

For an element in the subalgebra $\mathfrak{g}_{\geq 3}$,
\begin{equation*}
L=V\partial_x^2+\Phi\mathcal{D}\partial_x,
\end{equation*}
the expression \eqref{cu1} of the operator $\mathscr{P}_3(L)\nabla H$ means it taking value in $\mathfrak{g}_{\geq 3}$, and the other expression \eqref{cu2} implies it has the same order with $L$. Hence, the Poisson tensor defined by \eqref{cu1}\eqref{cu2} provides us with a Hamiltonian system corresponding to the isospectral flow generated by $L$.

The gradient $\nabla H$ is parameterized as
\begin{equation*}
\nabla H=-\partial_x^{-3}\mathcal{D}\frac{\delta H}{\delta V}+\partial_x^{-2}\frac{\delta H}{\delta \Phi}.
\end{equation*}
By direct calculation, we obtain
\begin{equation}\label{sys1}
\frac{\tt{d}}{{\tt d}t}\left(
\begin{array}{c}
V\\
\Phi
\end{array}\right)=2\left(
\begin{array}{cc}
P_{VV}&P_{V\Phi}\\
P_{\Phi V}&P_{\Phi\Phi}
\end{array}\right)\left(
\begin{array}{c}
\frac{\delta H}{\delta V}\\
\frac{\delta H}{\delta \Phi}
\end{array}\right)
\end{equation}
where
\begin{eqnarray*}
P_{VV}&=&-V^3\mathcal{D}^3-\mathcal{D}^3V^3+4(\mathcal{D}V)V_xV+\Phi V_xV-2\Phi V(\mathcal{D}\Phi)+(\mathcal{D}V)V(\mathcal{D}\Phi),\\
P_{V\Phi}&=&V^3\mathcal{D}^4+3\Phi V^2\mathcal{D}^3+\Big[2V_xV^2-V^2(\mathcal{D}\Phi)-3(\mathcal{D}V)\Phi V\Big]\mathcal{D}^2\\
&&+\Big[3\Phi_xV^2+\Phi V_xV+\Phi V(\mathcal{D}\Phi)\Big]\mathcal{D}+V_{xx}V^2-2V^2(\mathcal{D}\Phi_x)-2(\mathcal{D}V_x)\Phi V
-2(\mathcal{D}V)\Phi_x V, \\
P_{\Phi V}&=&-\mathcal{D}^4V^3+3\mathcal{D}^3\Phi V^2+\mathcal{D}^2\Big[2V_xV^2-V^2(\mathcal{D}\Phi)-3(\mathcal{D}V)\Phi V\Big]\\
&&-\mathcal{D}\Big[3\Phi_xV^2+\Phi V_xV+\Phi V(\mathcal{D}\Phi)\Big]-V_{xx}V^2+2V^2(\mathcal{D}\Phi_x)+2(\mathcal{D}V_x)\Phi V
+2(\mathcal{D}V)\Phi_x V, \\
P_{\Phi\Phi}&=&-\Phi_xV^2\mathcal{D}^2-\mathcal{D}^2\Phi_xV^2-2\Phi_x\Phi V\mathcal{D}-2\mathcal{D}\Phi_x\Phi V.
\end{eqnarray*}

Then under the transformation
\begin{equation*}\left(
\begin{array}{c}
V\\
\Phi
\end{array}\right)=\left(
\begin{array}{c}
U^2\\
(\mathcal{D}U)U+\Psi
\end{array}\right)
\end{equation*}
a modified system is obtained
\begin{equation}\label{sys2}
\frac{\tt{d}}{{\tt d}t}\left(
\begin{array}{c}
U\\
\Psi
\end{array}\right)=2\left(
\begin{array}{cc}
P_{UU}&P_{U\Psi}\\
P_{\Psi U}&P_{\Psi\Psi}
\end{array}\right)\left(
\begin{array}{c}
\frac{\delta H}{\delta U}\\
\frac{\delta H}{\delta \Psi}
\end{array}\right)
\end{equation}
with the Hamiltonian operator is given by
\begin{equation*}\left(
\begin{array}{cc}
P_{VV}&P_{V\Phi}\\
P_{\Phi V}&P_{\Phi\Phi}
\end{array}\right)_{V=U^2,\Phi=(\mathcal{D}U)U+\Psi}=\left(
\begin{array}{cc}
2U&0\\
\mathcal{D}\cdot U& 1
\end{array}
\right)
\left(
\begin{array}{cc}
P_{UU}&P_{U\Psi}\\
P_{\Psi U}&P_{\Psi\Psi}
\end{array}\right)\left(
\begin{array}{cc}
2U& -U\mathcal{D}\\
0&1
\end{array}
\right),
\end{equation*}
where
\begin{eqnarray*}
P_{UU}&=&-\frac{1}{2}\Big(U^2\mathcal{D}^3U^2+\Psi(\mathcal{D}\Psi)\Big)\\
P_{U\Psi}&=&\frac{1}{2}\Big[3\Psi V^3\mathcal{D}\partial_x-\Big(V^3(\mathcal{D}\Psi)+6(\mathcal{D}V)\Psi V^2\Big)\partial_x
+\Big(3\Psi_xV^3+3\Psi V_xV^2+(\mathcal{D}V)V^2(\mathcal{D}\Psi)\Big)\mathcal{D}\\
&&\quad-2V^3(\mathcal{D}\Psi_x)
-4(\mathcal{D}V_x)\Psi V^2-4(\mathcal{D}V)\Psi_xV^2-4(\mathcal{D}V)\Psi V_xV\Big]\\
P_{\Psi U}&=&\frac{1}{2}\Big[-3\Psi V^3\mathcal{D}\partial_x+\Big(2V^3(\mathcal{D}\Psi)+3(\mathcal{D}V)\Psi V^2\Big)\partial_x
+\Big((\mathcal{D}V)V^2(\mathcal{D}\Psi)-6\Psi V_xV^2\Big)\mathcal{D}\\
&&\quad+2V_xV^2(\mathcal{D}\Psi)
+V^3(\mathcal{D}\Psi_x)+4(\mathcal{D}V_x)\Psi V^2-(\mathcal{D}V)\Psi_xV^2
+4(\mathcal{D}V)\Psi V_xV\Big]\\
P_{\Psi\Psi}&=&\frac{1}{2}\Big[-V^6\mathcal{D}\partial_x^2-3(\mathcal{D}V)V^5\partial_x^2-6V_xV^5\mathcal{D}\partial_x
-\Big(4(\mathcal{D}V_x)V^5+13(\mathcal{D}V)V_xV^4+\Psi V^2(\mathcal{D}\Psi)\Big)\partial_x\\
&&\quad+\Big((\mathcal{D}V_x)(\mathcal{D}V)V^4-3V_{2x}V^5-6V_x^2V^4-V^2(\mathcal{D}\Psi)^2-9\Psi_x\Psi V^2-2(\mathcal{D}V)\Psi V(\mathcal{D}\Psi)\Big)\mathcal{D}\\
&&\quad+4\Psi_xV^2(\mathcal{D}\Psi)-4\Psi V^2(\mathcal{D}\Psi_x)-2(\mathcal{D}V_{2x})V^5-8(\mathcal{D}V)V_xV^4-6(\mathcal{D}V)V_{2x}V^4\\
&&\quad-8(\mathcal{D}V)V_x^2V^3
-8(\mathcal{D}V)\Psi_x\Psi V\Big].
\end{eqnarray*}

By setting $\Psi=0$, the system \eqref{sys2} is reduced to only one field whose Hamiltonian structure is obtained by Dirac reduction
\begin{eqnarray*}
P_{UU}\mid_{\Psi=0}-P_{U\Psi}\mid_{\Psi=0}\cdot(P_{\Psi\Psi}\mid_{\Psi=0})^{-1}\cdot P_{\Psi U}\mid_{\Psi=0}=-\frac{1}{2}\Big(U^2\mathcal{D}^3U^2\Big).
\end{eqnarray*}

\section{Proof of the Proposition \ref{prop2}:}\label{app4}
 Let's consider the first term in the l.h.s. of \eqref{jac},
\begin{eqnarray*}
&&\langle\alpha,Q^{\prime}_{Q(\beta)}(\gamma)\rangle\\
&=&\int {\tt d}x{\tt d}\theta\;\frac{3}{2}\alpha\Big(U^{1/2}Q(\beta)\mathcal{D}U^{1/2}\mathcal{D}U^{-1}\mathcal{D}^{-5}
U^{-1}\mathcal{D}U^{1/2}\mathcal{D}U^{3/2}\gamma\Big)\\
&&+\int {\tt d}x{\tt d}\theta\;\frac{1}{2}\alpha\Big(U^{3/2}\mathcal{D}U^{-1/2}Q(\beta)\mathcal{D}U^{-1}\mathcal{D}^{-5}
U^{-1}\mathcal{D}U^{1/2}\mathcal{D}U^{3/2}\gamma\Big)\\
&&-\int {\tt d}x{\tt d}\theta\;\alpha\Big(U^{3/2}\mathcal{D}U^{1/2}\mathcal{D}U^{-2}Q(\beta)\mathcal{D}^{-5}
U^{-1}\mathcal{D}U^{1/2}\mathcal{D}U^{3/2}\gamma\Big)\\
&&-\int {\tt d}x{\tt d}\theta\;\alpha\Big(U^{3/2}\mathcal{D}U^{1/2}\mathcal{D}U^{-1}\mathcal{D}^{-5}
U^{-2}Q(\beta)\mathcal{D}U^{1/2}\mathcal{D}U^{3/2}\gamma\Big)\\
&&+\int {\tt d}x{\tt d}\theta\;\frac{1}{2}\alpha\Big(U^{3/2}\mathcal{D}U^{1/2}\mathcal{D}U^{-1}\mathcal{D}^{-5}
U^{-1}\mathcal{D}U^{-1/2}Q(\beta)\mathcal{D}U^{3/2}\gamma\Big)\\
&&+\int {\tt d}x{\tt d}\theta\;\frac{3}{2}\alpha\Big(U^{3/2}\mathcal{D}U^{1/2}\mathcal{D}U^{-1}\mathcal{D}^{-5}
U^{-1}\mathcal{D}U^{1/2}\mathcal{D}U^{1/2}Q(\beta)\gamma\Big).
\end{eqnarray*}
Integrating all six terms by parts, we have
\begin{eqnarray*}
&&\langle\alpha,Q^{\prime}_{Q(\beta)}(\gamma)\rangle\\
&=&\int {\tt d}x{\tt d}\theta\;\frac{3}{2}U^{-1}\alpha Q(\beta)Q(\gamma)-\int {\tt d}x{\tt d}\theta\;\frac{3}{2}U^{-1}Q(\alpha)Q(\beta)\gamma\\
&&+\int {\tt d}x{\tt d}\theta\;U^{1/2}[\mathcal{D}^5A][\mathcal{D}U^{1/2}\mathcal{D}U^{-1}B]C-\int {\tt d}x{\tt d}\theta\;U^{1/2}A[\mathcal{D}U^{1/2}\mathcal{D}U^{-1}B][\mathcal{D}^5C]\\
&&+\int {\tt d}x{\tt d}\theta\;\frac{1}{2}U^{1/2}[\mathcal{D}^{-1}U\mathcal{D}^5A][\mathcal{D}U^{1/2}\mathcal{D}U^{-1}B][\mathcal{D}U^{-1}C]\\
&&-\int {\tt d}x{\tt d}\theta\;\frac{1}{2}U^{1/2}[\mathcal{D}U^{-1}A][\mathcal{D}U^{1/2}\mathcal{D}U^{-1}B][\mathcal{D}^{-1}U\mathcal{D}^5C],
\end{eqnarray*}
where
\begin{eqnarray*}
A&\equiv&\mathcal{D}^{-5}U^{-1}\mathcal{D}U^{1/2}\mathcal{D}U^{3/2}\alpha\\
B&\equiv&\mathcal{D}^{-5}U^{-1}\mathcal{D}U^{1/2}\mathcal{D}U^{3/2}\beta\\
C&\equiv&\mathcal{D}^{-5}U^{-1}\mathcal{D}U^{1/2}\mathcal{D}U^{3/2}\gamma.
\end{eqnarray*}
Therefore,
\begin{eqnarray*}
&&\langle\alpha,P^{\prime}_{P(\beta)}(\gamma)\rangle+\langle\beta,P^{\prime}_{P(\gamma)}(\alpha)\rangle
+\langle\gamma,P^{\prime}_{P(\alpha)}(\beta)\rangle\\
&=&\int {\tt d}x{\tt d}\theta\;\frac{3}{2}U^{-1}\alpha Q(\beta)Q(\gamma)-\int {\tt d}x{\tt d}\theta\;\frac{3}{2}U^{-1}Q(\alpha)Q(\beta)\gamma\\
&&+\int {\tt d}x{\tt d}\theta\;U^{1/2}[\mathcal{D}^5A][\mathcal{D}U^{1/2}\mathcal{D}U^{-1}B]C-\int {\tt d}x{\tt d}\theta\;U^{1/2}A[\mathcal{D}U^{1/2}\mathcal{D}U^{-1}B][\mathcal{D}^5C]\\
&&+\int {\tt d}x{\tt d}\theta\;\frac{1}{2}U^{1/2}[\mathcal{D}^{-1}U\mathcal{D}^5A][\mathcal{D}U^{1/2}\mathcal{D}U^{-1}B][\mathcal{D}U^{-1}C]\\
&&-\int {\tt d}x{\tt d}\theta\;\frac{1}{2}U^{1/2}[\mathcal{D}U^{-1}A][\mathcal{D}U^{1/2}\mathcal{D}U^{-1}B][\mathcal{D}^{-1}U\mathcal{D}^5C]\\
&&+\mbox{cyclic permutation of}(\alpha,\beta,\gamma)\\
&=&\int {\tt d}x{\tt d}\theta\;U^{1/2}[\mathcal{D}^5A][\mathcal{D}U^{1/2}\mathcal{D}U^{-1}B]C
-\int {\tt d}x{\tt d}\theta\;U^{1/2}A[\mathcal{D}U^{1/2}\mathcal{D}U^{-1}B][\mathcal{D}^5C]\\
&&+\underline{\int {\tt d}x{\tt d}\theta\;\frac{1}{2}U^{1/2}[\mathcal{D}^{-1}U\mathcal{D}^5A][\mathcal{D}U^{1/2}\mathcal{D}U^{-1}B][\mathcal{D}U^{-1}C]}\\
&&-\int {\tt d}x{\tt d}\theta\;\frac{1}{2}U^{1/2}[\mathcal{D}U^{-1}A][\mathcal{D}U^{1/2}\mathcal{D}U^{-1}B][\mathcal{D}^{-1}U\mathcal{D}^5C]\\
&&+\int {\tt d}x{\tt d}\theta\;U^{1/2}[\mathcal{D}^5B][\mathcal{D}U^{1/2}\mathcal{D}U^{-1}C]A
-\int {\tt d}x{\tt d}\theta\;U^{1/2}B[\mathcal{D}U^{1/2}\mathcal{D}U^{-1}C][\mathcal{D}^5A]\\
&&+\int {\tt d}x{\tt d}\theta\;\frac{1}{2}U^{1/2}[\mathcal{D}^{-1}U\mathcal{D}^5B][\mathcal{D}U^{1/2}\mathcal{D}U^{-1}C][\mathcal{D}U^{-1}A]\\
&&-\underline{\int {\tt d}x{\tt d}\theta\;\frac{1}{2}U^{1/2}[\mathcal{D}U^{-1}B][\mathcal{D}U^{1/2}\mathcal{D}U^{-1}C][\mathcal{D}^{-1}U\mathcal{D}^5A]}\\
&&+\int {\tt d}x{\tt d}\theta\;U^{1/2}[\mathcal{D}^5C][\mathcal{D}U^{1/2}\mathcal{D}U^{-1}A]B
-\int {\tt d}x{\tt d}\theta\;U^{1/2}C[\mathcal{D}U^{1/2}\mathcal{D}U^{-1}A][\mathcal{D}^5B]\\
&&+\int {\tt d}x{\tt d}\theta\;\frac{1}{2}U^{1/2}[\mathcal{D}^{-1}U\mathcal{D}^5C][\mathcal{D}U^{1/2}\mathcal{D}U^{-1}A][\mathcal{D}U^{-1}B]\\
&&-\int {\tt d}x{\tt d}\theta\;\frac{1}{2}U^{1/2}[\mathcal{D}U^{-1}C][\mathcal{D}U^{1/2}\mathcal{D}U^{-1}A][\mathcal{D}^{-1}U\mathcal{D}^5B].
\end{eqnarray*}
The two terms underlined can be integrated by parts, i.e.
\begin{eqnarray*}
&&\int {\tt d}x{\tt d}\theta\;\frac{1}{2}U^{1/2}[\mathcal{D}^{-1}U\mathcal{D}^5A]\Big([\mathcal{D}U^{1/2}\mathcal{D}U^{-1}B][\mathcal{D}U^{-1}C]
-[\mathcal{D}U^{-1}B][\mathcal{D}U^{1/2}\mathcal{D}U^{-1}C]\Big)\\
&=&\int {\tt d}x{\tt d}\theta\;\frac{1}{2}[\mathcal{D}^{-1}U\mathcal{D}^5A]\mathcal{D}\Big(-U^{-1}(\mathcal{D}C)(\mathcal{D}B)
+U^{-2}(\mathcal{D}U)(\mathcal{D}B)C-U^{-2}(\mathcal{D}U)(\mathcal{D}C)B\Big)\\
&=&-\int {\tt d}x{\tt d}\theta\;\frac{1}{2}[U\mathcal{D}^5A]\Big(-U^{-1}(\mathcal{D}C)(\mathcal{D}B)
+U^{-2}(\mathcal{D}U)(\mathcal{D}B)C-U^{-2}(\mathcal{D}U)(\mathcal{D}C)B\Big)\\
&=&\int {\tt d}x{\tt d}\theta\;\frac{1}{2}[\mathcal{D}^5A]\Big((\mathcal{D}C)(\mathcal{D}B)
-U^{-1}(\mathcal{D}U)(\mathcal{D}B)C+U^{-1}(\mathcal{D}U)(\mathcal{D}C)B\Big).
\end{eqnarray*}
Other terms involving nonlocal factor could be treated similarly. Finally, we have
\begin{eqnarray*}
&&\langle\alpha,P^{\prime}_{P(\beta)}(\gamma)\rangle+\langle\beta,P^{\prime}_{P(\gamma)}(\alpha)\rangle
+\langle\gamma,P^{\prime}_{P(\alpha)}(\beta)\rangle\\
&=&\int {\tt d}x{\tt d}\theta\;\Big(-(\mathcal{D}A_{2x})C_xB+(\mathcal{D}A_{2x})CB_x+(\mathcal{D}B_{2x})C_xA-(\mathcal{D}B_{2x})CA_x
-(\mathcal{D}C_{2x})B_xA\\
&&+(\mathcal{D}C_{2x})BA_x+{1\over2}(\mathcal{D}C_{2x})(\mathcal{D}B)(\mathcal{D}A)
+{1\over2}(\mathcal{D}C)(\mathcal{D}B_{2x})(\mathcal{D}A)
+{1\over2}(\mathcal{D}C)(\mathcal{D}B)(\mathcal{D}A_{2x})\Big)\\
&=&\int {\tt d}x{\tt d}\theta\;\mathcal{D}\Big(-C_{2x}B_xA+C_{2x}BA_x+C_xB_{2x}A-C_xBA_{2x}-CB_{2x}A_x+CB_xA_{2x}\\
&&+(\mathcal{D}B_x)(\mathcal{D}A_x)C-{1\over2}(\mathcal{D}B_x)(\mathcal{D}A)C_x-{1\over2}(\mathcal{D}B)(\mathcal{D}A_x)C_x
+{1\over2}(\mathcal{D}B)(\mathcal{D}A)C_{2x}\\
&&-(\mathcal{D}C_x)(\mathcal{D}A_x)B+{1\over2}(\mathcal{D}C_x)(\mathcal{D}A)B_x
+(\mathcal{D}C_x)(\mathcal{D}B_x)A-{1\over2}(\mathcal{D}C_x)(\mathcal{D}B)A_x\\
&&+{1\over2}(\mathcal{D}C)(\mathcal{D}A_x)B_x-{1\over2}(\mathcal{D}C)(\mathcal{D}A)B_{2x}
-{1\over2}(\mathcal{D}C)(\mathcal{D}B_x)A_x+{1\over2}(\mathcal{D}C)(\mathcal{D}B)A_{2x}\Big)\\
&=&0.
\end{eqnarray*}
The Jacobi identity holds.

\end{document}